\documentclass[aps,10pt,prc,tightenlines,twocolumn,twoside,showpacs,nofootinbib,superscriptaddress]{revtex4-2}

\usepackage{dblfloatfix}
\usepackage{blindtext}
\usepackage{ulem} %
\usepackage{ragged2e}

\usepackage{amsmath,amssymb,amsthm,slashed}   
\usepackage{hyperref}  
\usepackage{graphicx,color}
\usepackage{verbatim}  
\usepackage{tabularx,multirow,makecell} 
\usepackage{soul}   
\usepackage{siunitx}  
\usepackage{bbm} 

\def\be{\begin{eqnarray}}
\def\ee{\end{eqnarray}}

\begin{document}

\title{Transverse momentum structure of strange and charmed baryons: a light-front Hamiltonian approach}

\author{Zhimin Zhu}
\email{zhuzhimin@impcas.ac.cn}
\affiliation{Institute of Modern Physics, Chinese Academy of Sciences, Lanzhou, Gansu, 730000, China}
\affiliation{School of Nuclear Physics, University of Chinese Academy of Sciences, Beijing, 100049, China}
\affiliation{CAS Key Laboratory of High Precision Nuclear Spectroscopy, Institute of Modern Physics, Chinese Academy of Sciences, Lanzhou 730000, China}

\author{Tiancai~Peng}
\email{pengtc20@lzu.edu.cn} 
\affiliation{School of Physical Science and Technology, Lanzhou University, Lanzhou 730000, China}
\affiliation{Research Center for Hadron and CSR Physics, Lanzhou University and Institute of Modern Physics of CAS, Lanzhou 730000, China}

\author{Zhi Hu}
\email{huzhi0826@gmail.com}
\affiliation{Institute of Modern Physics, Chinese Academy of Sciences, Lanzhou, Gansu, 730000, China}
\affiliation{School of Nuclear Physics, University of Chinese Academy of Sciences, Beijing, 100049, China}
\affiliation{CAS Key Laboratory of High Precision Nuclear Spectroscopy, Institute of Modern Physics, Chinese Academy of Sciences, Lanzhou 730000, China}

\author{Siqi Xu}
\email{xsq234@impcas.ac.cn}
\affiliation{Institute of Modern Physics, Chinese Academy of Sciences, Lanzhou, Gansu, 730000, China}
\affiliation{School of Nuclear Physics, University of Chinese Academy of Sciences, Beijing, 100049, China}
\affiliation{CAS Key Laboratory of High Precision Nuclear Spectroscopy, Institute of Modern Physics, Chinese Academy of Sciences, Lanzhou 730000, China}

\author{Chandan Mondal}
\email{mondal@impcas.ac.cn}
\affiliation{Institute of Modern Physics, Chinese Academy of Sciences, Lanzhou, Gansu, 730000, China}
\affiliation{School of Nuclear Physics, University of Chinese Academy of Sciences, Beijing, 100049, China}
\affiliation{CAS Key Laboratory of High Precision Nuclear Spectroscopy, Institute of Modern Physics, Chinese Academy of Sciences, Lanzhou 730000, China}

\author{Xingbo Zhao}
\email{xbzhao@impcas.ac.cn}
\affiliation{Institute of Modern Physics, Chinese Academy of Sciences, Lanzhou, Gansu, 730000, China}
\affiliation{School of Nuclear Physics, University of Chinese Academy of Sciences, Beijing, 100049, China}
\affiliation{CAS Key Laboratory of High Precision Nuclear Spectroscopy, Institute of Modern Physics, Chinese Academy of Sciences, Lanzhou 730000, China}

\author{James P. Vary}
\email{jvary@iastate.edu}
\affiliation{Department of Physics and Astronomy, Iowa State University, Ames, IA 50011, USA}

\collaboration{BLFQ Collaboration}

\date{\today}

\begin{abstract}
Under the basis light-front quantization framework, we investigate the leading-twist transverse-momentum-dependent parton distribution functions (TMDs) for $\Lambda$ and $\Lambda_c$ baryons, the spin-1/2 composite systems consisting of two light quarks ($u$ and $d$) and a $s/c$ quark. We evaluate the TMDs using the overlaps of the light-front wave functions in the leading Fock sector, which are obtained by solving the light-front eigenvalue equation. We also study the spin densities of quarks in momentum space for various polarizations. In the same model, we compare the TMDs of the strange and charmed baryons and the proton by reviewing their spin structures in the quark model and the probabilistic interpretations of their TMDs. 
\end{abstract}
\maketitle

\section{Introduction}
Basis light-front quantization (BLFQ) has emerged recently as a promising nonperturbative tool to obtain the particle properties and observables from a Hamiltonian based, in part, on QCD.~\cite{Vary:2009gt,Honkanen:2010rc,Zhao:2013cma,Maris:2013qma,Wiecki:2014ola,Li:2017mlw,Lan:2019vui,Lan:2021wok,Mondal:2019jdg,Mondal:2021wfq,Xu:2021wwj,Peng:2022lte,Xu:2022abw}. 
By employing the light-front wave functions (LFWFs) of the strange and charmed baryons recently obtained within the BLFQ framework~\cite{Peng:2022lte}, we calculate the transverse-momentum-dependent parton distributions (TMDs) of the $ \Lambda $ and $ \Lambda_c $ baryons. Due to their short lifetime, it is difficult to directly extract the TMDs of $ \Lambda $ and $ \Lambda_c $ from experiments, but we provide additional motivations for conducting such calculations.

First, $ \Lambda $ is one of the central objects of current hypernuclear physics and its properties are also connected with the properties of neutron stars~\cite{Hashimoto:2006aw,Feliciello:2015dua,Vidana:2018bdi,Tolos:2020aln}. The study of the internal structure of the $ \Lambda $ baryons could also help us to understand the internal structure of hypernuclei more generally. Further, as the lightest charmed baryon, the $ \Lambda_c $ provides an experimental and theoretical place for studying the dynamics of light quark systems in a heavy quark environment, and for studying $CP$ violation in weak decays~\cite{ALEPH:1999syy,FOCUS:2005vxq,BESIII:2018ciw}.


Second, TMDs contain information about three-dimensional (3-D) structures including spin-momentum correlations inside the hadron~\cite{Bacchetta:2016ccz,Jaffe:1983hp,Collins:2011zzd}, so they are central objects of future EIC~\cite{Accardi:2012qut} and EicC experiments~\cite{Anderle:2021wcy}. There exist many theoretical calculations on the electromagnetic form factors of heavy baryons~\cite{Kubis:2000aa,Puglia:2000jy,Lin:2008mr,VanCauteren:2003hn,Julia-Diaz:2004yqv,Yang:2017hao,Haidenbauer:2016won,Dalkarov:2009yf,Faldt:2017kgy,Faldt:2016qee,Baldini:2007qg}. However, currently there exist few theoretical predictions of the strange and charmed baryon TMDs. Thus, our predictions can provide a baseline for future theoretical and experimental investigations of the 3-D structures of the strange and charmed baryons.

In this study, we diagonalize an effective Hamiltonian to obtain the LFWFs of $ \Lambda $ and $ \Lambda_c $, the overlaps of which give the TMDs. Currently, we truncate the Fock sector expansion to the valence Fock sector, which means that the proton, $ \Lambda $, and $ \Lambda_c $ are modeled as a bound state of $ uud $, $ uds $, and $ udc $ quarks, respectively. With this picture in mind, it is very interesting to test the influence of the quark mass by comparing the TMDs of those three baryons. We find consistency in expected mass effects within the qualitative and quantitative behaviors of the TMDs. In turn, this supports the use of the BLFQ framework to describe key properties of the heavy baryons.

The paper is organized as follows: in Sec.~\ref{Sec2} we introduce the light-front Hamiltonian approach under the BLFQ framework. In Sec.~\ref{Sec3}, we define the twist-2 TMDs of spin-1/2 baryons, and derive the overlap forms of the TMDs. Then we review their probabilistic interpretations. In Sec.~\ref{Sec4}, we present the numerical results of $\Lambda$ and $\Lambda_c$ baryons and compare them with those of the proton. Finally, we summarize our work in Sec.~\ref{Sec5}.
\section{BASIS LIGHT-FRONT QUANTIZATION\label{Sec2}}

\subsection{A light-front Hamiltonian approach}
In light-front field theory~\cite{Kogut:1969xa}, the light-front variables are defined as $V^\pm\equiv V^0\pm V^3$, $\vec V_\perp\equiv(V_1,V_2)$\footnote{Here we follow a different convention than Refs.~\cite{Kogut:1969xa,Barone:2001sp}. Thus, the projection operators and the phase factor in the definitions of TMD correlators in Eq.~(\ref{TMD}) are also different.}, and the energy-momentum relation is $P^+P^--P^2_\perp=M^2$. Upon quantization, this provides the light-front eigenvalue equation 
\begin{equation}
    (P^+P^--P^2_\perp)|P,\Lambda\rangle=M^2|P,\Lambda\rangle,
    \label{SchrodingerEq}
\end{equation}
where $|P,\Lambda\rangle$ is the light-front state with the momentum $P$ and the light-front helicity $\Lambda$. $M$ is the system mass. We will focus on a bound state solution which is expanded in the Fock space as~\cite{Brodsky:1997de}
\begin{align}
    |P,\Lambda\rangle&=\sum_n\sum_{\lambda_{1},\lambda_{2},\cdots,\lambda_{n}} \int \prod_{i}^{n}\frac{\left[\mathrm{d} x_{i} \mathrm{d}^2 \vec k_{\perp i}\right]}{2(2\pi)^3\sqrt{x_i}}\nonumber\\
    &\times2(2\pi)^3 \delta\left(1-\sum_{i}^{n} x_{i}\right)\delta^{2}\left(\vec P_{\perp}-\sum_{i}^{n} \vec k_{\perp i}\right) \nonumber\\
    &\times\Psi_{n,\{\lambda_{i}\}}^{\Lambda}\left(\left\{x_i, k_{\perp i}\right\}\right)\left|\{x_i P^+,\vec k_{\perp i} +x_i \vec P_{\perp},\lambda_i\}\right\rangle ,
    \label{BoundState0}
\end{align}
where $i$ is the index of the parton inside the bound state, $x_i\equiv k^+_i/P^+$ refers to the longitudinal momentum fraction, $k^+_i$ is the longitudinal momentum of the parton, $\vec k_{\perp i}$ is the intrinsic transverse momentum, and $\lambda_i$ is the light-front helicity. The above two $\delta$ functions ensure the conservation of momentum in the longitudinal direction and the transverse plane. The LFWF, $\Psi^\Lambda_{n,\{\lambda_i\}}$, is boost invariant and independent of the hadron momentum, $(P^+, \vec P_\perp)$, but depends on the parton momenta, $\{x_i,\vec k_{\perp i}\}$, and the parton helicities $\{\lambda_i\}$. The Fock state $|\psi_a\rangle$ is 
\begin{equation}
    \begin{aligned}
        |\psi_a\rangle\equiv |\{x_i P^+,\vec k_{\perp i} +x_i \vec P_{\perp},\lambda_i\}\rangle=\hat b_{i^{\prime}}^\dagger \cdots\hat  d_{j^{\prime}}^\dagger \cdots\hat  a_{k^{\prime}}^\dagger \cdots|0\rangle,
    \end{aligned}
\end{equation}
where $\hat b^\dagger$, $\hat d^\dagger$, and $\hat a^\dagger$ represent the creation operators of quarks, antiquarks, and gluons, respectively. $|0\rangle$ is the light-front vacuum state.

From Eq.~(\ref{SchrodingerEq}), the light-front Hamiltonian matrix element is expressed as
\begin{equation}
    H_{ab}=\langle \psi_a|(P^+P^--P^2_\perp)| \psi_b\rangle.
    \label{the_light_front_matrix}
\end{equation}
The eigenequation, Eq.~(\ref{SchrodingerEq}), can then be converted to the following matrix form 
\begin{equation}
    H_{ab}\Psi_b=M^2\Psi_a
    \label{the_light_front_Schrodinger_equation},
\end{equation}
which is solved to obtain the Fock sector related LFWFs $\{\Psi_a\}$ that encode the structural information of the bound states.

At a fixed light-front time, $x^+\equiv x^0+x^3$, the bound state of a baryon can be expressed in terms of $|qqq\rangle$, $|qqqq\bar q\rangle$, $|qqqg\rangle$, and other Fock sectors~\cite{Xu:2021wwj}. For numerical calculations, we must truncate the infinite Fock sector expansion to a finite Fock space in Eq.~(\ref{BoundState0}). In this work, the baryon bound states are restricted to the valence Fock sector, which means there are only the three-quark LFWFs $\Psi^\Lambda$ in Eq.~(\ref{BoundState0}). 
Instead of the full light-front QCD Hamiltonian, $H_{\rm QCD}\equiv P^+P^-_{\rm QCD}-P^2_\perp$, the current Hamiltonian we use, $H$, contains an effective Hamiltonian ${H}_{\text{eff}}$ and a constraint term $H^\prime$~\cite{Mondal:2019jdg,Mondal:2021wfq,Xu:2021wwj,Peng:2022lte},
\begin{equation}
    H=H_{\text{eff}}+H^\prime.
    \label{Hamiltonian1}
\end{equation}

For the valence Fock sector of the baryon, the effective Hamiltonian consists of the kinetic energy of quarks, a confining potential, and the one-gluon exchange (OGE) interaction~\cite{Mondal:2019jdg,Mondal:2021wfq,Xu:2021wwj,Peng:2022lte},
\begin{equation}
    H_{\mathrm{eff}}=\sum_{i} \frac{\vec{k}_{\perp i}^{2}+m_{i}^{2}}{x_{i}}+\frac{1}{2} \sum_{i, j} V_{i, j}^{\mathrm{conf}}+\frac{1}{2} \sum_{i, j} V_{i, j}^{\mathrm{OGE}},
    \label{Hamiltonian}
\end{equation}
where the longitudinal momentum fraction is conserved $\sum_i x_i =1$. 
For compactness of notation, we will define these terms in mixed coordinate and momentum space variables where there is no ordering ambiguity. Ultimately, they will be evaluated in a BLFQ basis space with integrations over all coordinates and momenta.
We adopt a confining potential which contains the transverse and the longitudinal parts as employed in Refs.~\cite{Xu:2021wwj,Mondal:2021wfq,Li:2015zda}
\begin{equation}
        V_{i, j}^{\mathrm{conf}}=\kappa^{4}\vec r_{ \perp ij}^{2}-\frac{\kappa^{4}}{\left(m_{i}+m_{j}\right)^{2}}\partial_{x_{i}}\left(x_{i} x_{j} \partial_{x_{j}}\right),
\end{equation}
where $\vec r_{\perp  ij}=\sqrt{x_i x_j}(\vec r_{\perp i}-\vec r_{\perp j})$ signifies the relative coordinate.  $\kappa$ represents the strength of the confinement, and $\partial_x \equiv (\partial/\partial x)_{r_{ \perp ij}}$.

The last term in Eq.~(\ref{Hamiltonian}) represents the OGE potential
\begin{equation}
    V_{i, j}^{\mathrm{OGE}}=\frac{ 4 \pi C_{F}\alpha_{s}}{Q_{ij}^{2}} \bar{u}_{s_{i}^{\prime}}\left(k_{i}^{\prime}\right) \gamma^{\mu} u_{s_{i}}(k_{i}) \bar{u}_{s_{j}^{\prime}}\left(k_{j}^{\prime}\right) \gamma_{\mu} u_{s_{j}}\left(k_{j}\right),
\end{equation}
where $C_F=-2/3$ is the color factor, and $\alpha_s$ is the coupling constant. $\bar u_{s_{i}}(k_i)$ and $u_{s_{i}}(k_i)$ represent the spinor wave functions. $Q^2_{ij}$ is the kinematical variable,
\begin{align}
    Q_{ij}^{2}=& \frac{1}{2}\left[\left(\frac{\vec{p}_{\perp i}^{2}+m_{i}^{2}}{x_{i}}-\frac{\vec{p}_{\perp i}^{\prime 2}+m_{i}^{2}}{x_{i}^{\prime}}\right.\right.\nonumber\\
    &\left.\left.-\frac{\left(\vec{p}_{\perp i}^{2}-\vec{p}_{\perp i}^{\prime 2}\right)+\mu_{g}^{2}}{x_{i}-x_{i}^{\prime}}\right)-(i \rightarrow j)\right],
    \label{Q2ij}
\end{align}
where $\mu_g$ is the gluon mass that regulates the infrared divergence in OGE.

Since we will be working in an overcomplete basis (see the next subsection for details) we require a Lagrange multiplier term to isolate the internal motion from the spurious center-of-mass (c.m.) motion in the LFWFs. Therefore in Eq.~(\ref{Hamiltonian1}), we introduce the constraint term
\begin{equation}
    H^\prime=\lambda_L\left(H_{\text{c.m.}}-2b^2I\right),
    \label{H_core}
\end{equation}
which effectively drives a factorization of the transverse c.m. motion from the intrinsic motion, where $2b^2$ is the two-dimensional harmonic oscillator (2-D HO) zero-point energy {(see below)}, and $\lambda_L$ is a Lagrange multiplier~\cite{wiecki2015basis,Xu:2021wwj,Mondal:2021wfq}. The c.m. motion is governed by
\begin{equation}
    H_{\text{c.m.}}=\left(\sum_i \vec k_{\perp i}\right)^2+b^4\left(\sum_i x_i \vec r_{\perp i}\right)^2,
    \label{Hamiltonian_c_m}
\end{equation}
where $\vec r_{\perp i}$ is the coordinate of each quark. One can set $\lambda_L$ sufficiently large to shift the excited states of the c.m. motion to higher energy away from the low-lying states~\cite{wiecki2015basis,Xu:2021wwj,Mondal:2021wfq}.

According to Eqs.~(\ref{Hamiltonian})-(\ref{Q2ij}), four model parameters will be introduced to solve the light-front eigenvalue equation, Eq.~(\ref{the_light_front_Schrodinger_equation}). Those parameters are: the coupling constant, the kinetic/OGE masses, and the strength of the confining potential.

\subsection{The BLFQ framework}
In this section, we introduce BLFQ, which provides a computational framework for solving the relativistic many body bound state problem in quantum field theories~\cite{Vary:2009gt,Honkanen:2010rc,Zhao:2013cma,Maris:2013qma,Wiecki:2014ola,Li:2017mlw,Lan:2019vui,Lan:2021wok,Mondal:2019jdg,Mondal:2021wfq,Xu:2021wwj,Peng:2022lte,Xu:2022abw}. To solve the eigenvalue equation, Eq.~(\ref{the_light_front_Schrodinger_equation}), we first calculate the light-front Hamiltonian matrix, Eq.~(\ref{the_light_front_matrix}), in a chosen basis space. The Fock-sector basis states in Eq.~(\ref{the_light_front_matrix}) are taken to be direct products of the single-particle states $|\alpha\rangle=\otimes_i|\alpha_i\rangle$. For simplicity, we take every single-particle basis state $|\alpha_i\rangle$ to be the direct product of the momentum eigenstates in the longitudinal direction, the 2-D HO basis states in the transverse plane, and the light-cone helicity eigenstates.

In the longitudinal direction, we adopt the Discretized Light Cone Quantization (DLCQ) basis \cite{Brodsky:1997de} using the standard normalization in a one-dimensional box of length $2L$,
\begin{equation}
    k_i^+=\frac{2\pi}{L}k_i,
\end{equation}
where $k_i$ is an integer (half-integer) for bosons (fermions). In the transverse plane, we use the 2-D HO basis function given by
\begin{align}
    \phi^m_{n}\left(\vec{k}_{\perp} ; b\right)=& \frac{1}{b} \sqrt{\frac{4\pi\times n !}{(n+|m|) !}} e^{-\vec{k}_{\perp}^{2} /\left(2 b^{2}\right)} \nonumber\\
    & \times\left(\frac{|\vec{k}_{\perp}|}{b}\right)^{|m|} L_{n}^{|m|}\left(\frac{\vec{k}_{\perp}^{2}}{b^{2}}\right) e^{i m \theta},
\end{align}
where $\theta=\arg{}(\vec k_\perp)$, $L^{|m|}_n$ is the associated Laguerre polynomial. The radial quantum number $n_i$, the orbital quantum number $m_i$, and the HO basis scale parameter $b$ define the HO energy $E_{n_i,m_i}=(2n_i+|m_i|+1)b^2$. Each single-particle basis state contains four quantum numbers $|\alpha_i\rangle=|k_i,n_i,m_i,\lambda_i\rangle$, where $\lambda_i$ refers to the light-front helicity. All the Fock-sector basis states have the same total angular momentum projection $\Lambda$ since it is conserved by our Hamiltonian, $H$, defined in Eq.~(\ref{Hamiltonian1}), 
\begin{equation}
    \sum_i(\lambda_i+m_i)=\Lambda.
\end{equation}
For the limited valence Fock space of the present work, we may suppress the flavor and color degrees of freedom.

For the purpose of numerical calculations, we must truncate the infinite basis of the leading (valence quarks only) Fock sector. In the longitudinal direction, the sum of the longitudinal momentum of all the basis states is the same as the longitudinal momentum of the Fock particles in the bound state $P^+=\sum_i k^+_i$. We then use a dimensionless variable $K=\sum_i k_i$ to parameterize $P^+$, and the longitudinal momentum fraction $x$ is defined as $x_i=k^+_i/P^+=k_i/K$. In the transverse plane, we truncate the total transverse quantum numbers such that
\begin{equation}
    N_{\text{max}}\geq\sum_i(2n_i+|m_i|+1).
    \label{Nmax}
\end{equation}
$N_{\text{max}}$ determines the ultraviolet ($\thicksim b\sqrt{N_\text{max}}$) cutoff and the infrared ($\thicksim b/\sqrt{N_\text{max}}$) cutoff in momentum space~\cite{Zhao:2014xaa}. 


After setting the truncation parameter $\{N_{\text{max}},K\}$ and solving the eigenvalue equation, Eq.~(\ref{the_light_front_Schrodinger_equation}), within the BLFQ framework, we obtain the LFWFs in the momentum space as
\begin{align}
    &\Psi^{\Lambda}_{\{ \lambda_{i}\}}({\{x_{i}, \vec{p}_{\perp i}\}}){=\langle P, \Lambda |\{x_{i}, \vec{p}_{\perp i}, \lambda_{i}\}\rangle} \nonumber\\
    &=\sum_{\{n_i,m_i\}}\Psi^{\Lambda}(\{x_{i}, n_{i}, m_{i}, \lambda_{i}\}) \prod_{i} \phi^{m_i}_{n_{i}}\left(\vec{p}_{\perp i} ; b\right),
\end{align}
where $\Psi^{\Lambda}(\{x_{i}, n_{i}, m_{i}, \lambda_{i}\})$ is the LFWF in the BLFQ basis obtained by diagonalizing Eq.~(\ref{the_light_front_Schrodinger_equation}). $\vec{p}_{\perp i}$ is the transverse  single-particle momentum.

For the LFWFs of the bound states in Eq.~(\ref{BoundState0}) and TMDs (see below), the transverse variable $ \vec k_\perp$ is the intrinsic transverse momentum. The BLFQ method adopts the 2-D HO basis states in the transverse plane which enables one to transform the single-particle coordinate into the relative coordinate by the Talmi-Moshinsky (TM) transform~\cite{Tobocman:1981yao} as 
\begin{equation}
    \begin{aligned}
        \phi^{m_1}_{n_{1}}(p_1)\phi^{m_2}_{n_{2}}(p_2)=\sum_{NMnm}M^{NMnm}_{n_1m_1n_2m_2}\phi^{M}_{N}(K)\phi^{m}_{n}(k),
    \end{aligned}\label{TMtransfer}
\end{equation}
where $M^{NMnm}_{n_1m_1n_2m_2}$ is the TM coefficient, the labels $(N,M)$ represent the c.m. quantum numbers, and $(n,m)$ represent the relative quantum numbers. 

For the baryons truncated to the valence quark Fock sector, when we need the LFWF expressed in terms of internal coordinates alone, the procedure for converting the single-particle coordinates to the relative is as follows: first, quark 1 and quark 2 are TM transformed to obtain the quantum numbers $(N_{12},M_{12})$ of their c.m. and the quantum numbers ($n_{12}$,$n_{12}$) of the relative motion via Eq.~(\ref{TMtransfer}). Second, these quantum numbers are TM transformed with quark 3, and finally, we get the quantum numbers $(n,m)$ of the struck quark 3 relating to the system center of mass. 

\subsection{\label{SpinStructure}The spin structure of baryons in the quark model}

In this subsection, we review the spin structure of the $\Lambda$ and $\Lambda_c$ baryon in the quark model. The results will be used to identify the $\Lambda$ and $\Lambda_c$ baryon states from BLFQ and analyze their structure.

In the quark model~\cite{Gell-Mann:1964ewy}, light and strange baryons are composed of three light or strange quark $qqq$. Despite the different masses of light and strange quarks, the $\rm{SU}(3)$ flavor symmetry holds approximately in nature. 
Therefore, we can still analyze the structures of light and strange baryons under the framework of $\rm{SU}(3)$ flavor symmetry.
In the flavor (f) $\rm{SU}(3)$ framework, ${q}^{\mathrm{f}}{q}^{\mathrm{f}}{q}^{\mathrm{f}}=3\otimes 3\otimes 3=10_s\oplus 8_\rho\oplus 8_\gamma\oplus 1_a$, subscript ($a$) $s$ represents the total (anti)symmetry, and subscripts $\rho$ and $\gamma$ refer to the mixed symmetry. In spin (s) space, ${q}^{\mathrm{s}}{q}^{\mathrm{s}}{q}^{\mathrm{s}}=2\otimes 2\otimes 2=4_s\oplus 2_\rho\oplus 2_\gamma$. In non-relativistic models, the quarks of the baryon ground states have zero orbital angular momentum, and the spatial wave function is symmetric (S wave).

Due to the Pauli exclusion principle and the color confinement, the flavor-spin wave functions of baryon ground states must be symmetric while the color wave function is the antisymmetric color singlet $1_a$. In the quark model, the $\Lambda$ baryon and the proton belong to the baryon octet. With the flavor-spin symmetry $\rm{SU}(6)$ analysis, the flavor-spin wave functions of the baryon octet are $\eta =\frac{1}{\sqrt{2}}(\phi^\rho \chi^\rho+\phi^\gamma \chi^\gamma)$, where $\phi$ is the flavor wave function, and $\chi$ is the spin wave function. For $\Lambda$ baryons with positive helicity, the flavor-spin wave function is 
\begin{widetext}
\begin{align}
        |\Lambda,\uparrow\rangle_{\text{flavor-spin}}&= \frac{1}{\sqrt{2}}\Bigg[\frac{1}{2}\Big(|sud\rangle+|usd\rangle-|sdu\rangle-|dsu\rangle\Big)\otimes\frac{1}{\sqrt{6}}\Big(|\uparrow\downarrow\uparrow\rangle+|\downarrow\uparrow\uparrow\rangle-2|\uparrow\uparrow\downarrow\rangle\Big)\nonumber\\
        &+\frac{1}{\sqrt{12}}\Big(|dsu\rangle-|sdu\rangle+|sud\rangle-|usd\rangle+2|uds\rangle-2|dus\rangle\Big)\otimes\frac{1}{\sqrt{2}}\Big(|\uparrow\downarrow\uparrow\rangle-|\downarrow\uparrow\uparrow\rangle\Big)\Bigg]. \label{Lambda_spin_flavor_wave_function1}
\end{align} 
\end{widetext}
Using the spin operator $\hat\sigma_{q}$ of quarks, the spin projection $\langle\hat\sigma_{q}\rangle=\langle\Lambda,\uparrow|\hat\sigma_{q}|\Lambda,\uparrow\rangle$ of both the $u$ quark and the $d$ quark inside $\Lambda$ baryons is zero, while that of the $s$ quark is +1/2, which means the $s$ quark is always parallel to the $\Lambda$ baryon. 
However, the proton has a different spin structure given by
\begin{widetext}
\begin{align}
        |p,\uparrow\rangle_{\text{flavor-spin}}=& \frac{1}{\sqrt{2}}\Bigg[\frac{1}{\sqrt{6}}\Big(2|uud\rangle-|duu\rangle-|udu\rangle\Big)\otimes\frac{1}{\sqrt{6}}\Big(|\uparrow\downarrow\uparrow\rangle+|\downarrow\uparrow\uparrow\rangle-2|\uparrow\uparrow\downarrow\rangle\Big)\nonumber\\
        &+\frac{1}{\sqrt{2}}\Big(|duu\rangle-|udu\rangle\Big)\otimes\frac{1}{\sqrt{2}}\Big(|\uparrow\downarrow\uparrow\rangle-|\downarrow\uparrow\uparrow\rangle\Big)\Bigg].        \label{proton_spin_flavor_wave_function11}
\end{align}
\end{widetext}
It shows $\langle\hat\sigma_{u}\rangle=2/3$ and $\langle\hat\sigma_{d}\rangle=-1/6$, which means that the proton spin at a low scale is primarily carried by the $u$ quarks.

In flavor space, $c$ quarks with a heavy mass will break flavor symmetry. The charmed baryon $\Lambda_c$ does not belong to the three light quark multiplet but to the two light quark system: $q^{\mathrm{f}} q^{\mathrm{f}}=3\otimes 3=\bar 3_a \oplus 6_s$ - the antitriplet $\bar 3_a$ \cite{Roberts:2007ni,Lu:2016ogy}.
For $\Lambda_c$, the flavor wave function is antisymmetric under the exchange of the first two quarks, $\phi=\frac{1}{\sqrt{2}}(|u d c \rangle-|d u c \rangle)$. The spin wave function must also be antisymmetric under the exchange of the first two quarks. So the flavor-spin wave function of $\Lambda_c$ is 
\begin{widetext}
    \begin{equation}
        |\Lambda_c,\uparrow\rangle_{\text{flavor-spin}}=\frac{1}{\sqrt{2}}\Big(|u d c \rangle-|d u c \rangle\Big)\otimes\frac{1}{\sqrt{2}}\Big(|\uparrow\downarrow\uparrow\rangle-|\downarrow\uparrow\uparrow\rangle\Big),
    \end{equation}
\end{widetext}
which shows the same spin structure as $\Lambda$. The spin projection of $u$ and $d$ quarks is zero, $\langle \hat{\sigma}_{u,d}\rangle=0$, and the $c$ quark is always parallel to the $\Lambda_c$ baryon, $\langle \hat{\sigma}_c\rangle=1/2$.

In conclusion, according to the quark model, $u$ and $d$ quarks have no contributions to the spin of the $\Lambda$ and $\Lambda_c$ baryon in S waves, only the $s$ ($c$) quark contributes to the spin of the $\Lambda$ ($\Lambda_c$) baryon. Both $u$ and $d$ quarks contribute to the proton spin, but the $u$ quarks dominate over the $d$ quark.

\section{TMDs\label{Sec3}}
Leading-twist TMDs provide the densities or differences of densities for a struck parton having the longitudinal momentum fraction $x$, relative transverse momentum $\vec k_{\perp}$, and a particular polarization in a hadron~\cite{Jaffe:1983hp,Collins:2011zzd}. For spin-1/2 baryons, TMDs of quarks are defined through the quark-quark correlator function as~\cite{Boer:1997nt,Bacchetta_2007,Meissner:2009ww},
\begin{align}
    \Phi_q^{[\Gamma]}&\left(P, S; x=\frac{k^{+}}{P^{+}}, \vec k_{\perp}\right)=\frac{1}{2} \int \frac{\mathrm{d} z^{-} \mathrm{d}^2 z_{\perp}}{2(2 \pi)^{3}} e^{i k \cdot z}\nonumber\\
    &\times\left.\left\langle P, S\right|{\bar{\psi}}_q(0) \Gamma \mathcal{W}(0_\perp,z_\perp) {\psi}_q(z)\left| P, S\right\rangle\right|_{z^{+}=0},\label{TMD}
\end{align}
where $\psi_q$ is the quark field operator, and index $q$ means a particular flavor. 
The quark fields in Eq.~(\ref{TMD}) are accompanied by the gauge links $\mathcal{W}$, necessary to render the gauge-invariance~\cite{Mulders:1995dh,Bacchetta:2008af}. $|P,S\rangle$ defines the bound state of a baryon with spin $S$ and four-momentum $P$ where the transverse momentum is zero $P_\perp=0$~\cite{Collins:1992kk}. The Dirac matrix $\Gamma$ determines the Lorentz structure of the correlator $\Phi^{[\Gamma]}$ and its `twist' $\tau$~\cite{Jaffe:1991kp}. For the leading twist ($\tau=2$), the Dirac matrices $\Gamma$ can only take three kinds, $\gamma^+$, $\gamma^+\gamma^5$, and $i\sigma^{i+}\gamma^5$ ($i=1,2$). In the light-front field theory~\cite{Kogut:1969xa}, the quark field is decomposed into a `good' component $\psi_+$, and a `bad' component $\psi_-$ ($\psi_{\pm}\equiv\frac{1}{4}\gamma^{\mp}\gamma^{\pm}\psi$), whereas the bad component $\psi_-$ is the dependent variable of the `good' component $\psi_+$ and the gauge fields,
\begin{equation}
    \psi_-(z)=\frac{\gamma^+}{2i\partial^+}[i(\partial_j-igA_j(z))\gamma_j+m]\psi_+(z),
\end{equation}
where the constraint equation comes from the QCD equation of motion in the light-cone gauge $A^+=0$. Fortunately, the leading-twist Dirac matrices project the correlator into terms containing only `good' fields, so the correlator has no additional complexity and no suppression in the power of $M/P^+$~\cite{Bacchetta:2019qkv}.

By analyzing parity, charge conjugation, hermiticity invariance, and using Gordon identities, one can parameterize the quark-quark correlator of spin $1/2$ baryons to get eight leading-twist TMDs~\cite{Boer:1997nt,Bacchetta_2007,Meissner:2009ww},
\begin{align}
    \Phi^{\left[\gamma^{+}\right]}&(x, \vec k_{\perp} ; S)=f_{1}-\frac{\epsilon_{\perp}^{i j} k_\perp^{i} S_\perp^{j}}{M} f_{1 T}^{\perp}, \label{TMDs1}\\
    \Phi^{\left[\gamma^{+} \gamma^{5}\right]}&(x, \vec k_{\perp} ; S)=S^{3} g_{1 L}+\frac{\vec k_{\perp} \cdot \vec S_\perp}{M} g_{1 T}, \label{TMDs2}\\
    \Phi^{\left[i \sigma^{j+} \gamma^{5}\right]}&(x, \vec k_{\perp} ; S)=S_\perp^{j} h_{1}+S^{3} \frac{k_\perp^{j}}{M} h_{1 L}^{\perp}\nonumber\\
    &+S_\perp^{i} \frac{2 k_\perp^{i} k_\perp^{j}-\left(\vec k_{\perp}\right)^{2} \delta^{i j}}{2 M^{2}} h_{1 T}^{\perp}+\frac{\epsilon_{\perp}^{j i} k_\perp^{i}}{M} h_{1}^{\perp},
    \label{TMDs3}
\end{align}
where $i,j=1,2$ and antisymmetric tensor $\epsilon^{12}_\perp=-\epsilon^{21}_\perp=1$. Based on Jaffe–Ji classification~\cite{Barone:2001sp}, the letters $f$, $g$, and $h$ respectively refer to unpolarized, longitudinally polarized, and transversely polarized struck quarks; subscript L (T) refers to the longitudinal (transverse) polarization of the baryon; subscript $1$ indicates the leading twist; the $\perp$ symbol represents a transverse momentum dependence with an uncontracted index. 

If one takes the naive time-reversal symmetry into account, the Sivers function $f^{\perp}_{1T}$~\cite{Sivers:1990fh} and the Boer-Mulders function $h^{\perp}_{1}$~\cite{Boer:1997nt} will disappear~\cite{Collins:1992kk,Tangerman:1994eh}. These functions are called T-odd functions. The rest are T-even TMDs~\cite{Meissner:2009ww}. The T-odd effect of TMDs was first mentioned in Ref.~\cite{Sivers:1990fh}. Reference~\cite{Boros:1993ps} provides an intuitive picture of quark orbital angular momentum and what they called `surface effect'.
If one wants to get the non-zero results of the T-odd functions in semi-inclusive deep inelastic scattering or Drell-Yan process, one must take the final or initial state interactions into account~\cite{Brodsky:2002rv}. In this present work, we do not consider the effect of the gauge links $\mathcal{W}$. In the unit matrix approximation, $\mathcal{W}\approx\mathbbm{1}$, only the T-even TMDs survive.

\subsection{Overlap representations \label{Overlap_representation}}

Based on light-front field theory~\cite{Kogut:1969xa}, we can obtain the TMDs as overlaps of the LFWFs. The critical step is to separate different TMDs in Eqs.~(\ref{TMDs1}-\ref{TMDs3}). We decompose the bound state of the baryon $|P,S\rangle$ in terms of the light-front helicity state of the baryon $|P,\Lambda\rangle$ by the rotation transformation~\cite{Lorce:2011zta},
\begin{equation}
    (|P,+S\rangle,|P,-S\rangle)=(|P,+\rangle,|P,-\rangle)u(\theta,\varphi),
\end{equation}
where the baryon polarization state $|P,S\rangle$ is in a generic direction $S=(\sin\theta \cos\varphi,\sin\theta \sin \varphi, \cos \theta)$. The $\rm{SU}(2)$ rotational matrix is 
\begin{equation}
    u(\theta,\varphi)=\left(\begin{array}{cc}
    \cos \frac{\theta}{2} e^{-i \frac{\varphi}{2}} & -\sin \frac{\theta}{2} e^{-i \frac{\varphi}{2}} \\
    \sin \frac{\theta}{2} e^{i \frac{\varphi}{2}} & \cos \frac{\theta}{2} e^{i \frac{\varphi}{2}}
    \end{array}\right).
\end{equation}
Therefore, one can represent the correlator, Eq.~(\ref{TMD}), in the light-front helicity form as follows, 
\begin{align}
    \Phi_{{\Lambda^\prime} \Lambda;q}^{[\Gamma]}( x, \vec k_{\perp})&=\left.\frac{1}{2} \int \frac{\mathrm{d} z^{-} \mathrm{d}^2 z^{\perp}}{2(2 \pi)^{3}} e^{i k \cdot z}\right.\nonumber\\
    &\times\left.\left\langle P, \Lambda^\prime\right|{\bar{\psi}}_q(0) \Gamma {\psi}_q(z)\left| P, \Lambda\right\rangle\right|_{z^{+}=0}.\label{TMDsdefinition}
 \end{align}

In the spinor space, the struck quarks have different helicity structures for different gamma matrices $\Gamma$ in Eq.~(\ref{TMDsdefinition}). We define the TMD correlators in terms of the helicity amplitude as
\begin{equation}
    \Phi_{{\Lambda^\prime} \Lambda;q}^{[\Gamma]}( x, \vec k_{\perp})=\sum_{\lambda\lambda^\prime}\Phi_{\Lambda^\prime\lambda^\prime,\Lambda\lambda;q}(x,\vec k_\perp)\bar{u}_{\lambda^\prime}(k)\Gamma u_\lambda(k).\label{overlap_form}
\end{equation}
Based on the light-cone bound states of the baryon in Eq.~(\ref{BoundState0}) and the anti-commutator of fermionic fields, we can obtain the overlap representation of the TMD correlators in Eq.~(\ref{overlap_form}),
\begin{align}
    \Phi&_{\Lambda^\prime\lambda^\prime,\Lambda\lambda;q}(x,\vec k_\perp)=\sum_{\lambda_2 \lambda_3}\int\frac{\mathrm{d}x_2\mathrm{d}^2\vec k_{\perp 2}\mathrm{d}x_3\mathrm{d}^2\vec k_{\perp 3}}{2(2\pi)^3(2\pi)^3}\nonumber\\
    &\times\delta(1-x-x_2-x_3)\delta^{(2)}(\vec k_\perp+\vec k_{\perp2}+\vec k_{\perp3})\nonumber\\
    &\times\Psi^{\Lambda^{\prime}*}_{\lambda^{\prime}\lambda_2\lambda_3}(\tilde{k},\tilde{k}_2,\tilde{k}_3)\Psi^{\Lambda}_{\lambda\lambda_2\lambda_3}(\tilde{k},\tilde{k}_2,\tilde{k}_3),\label{TMDsoverlap}
\end{align}
where $\tilde{k}\equiv(x,\vec k_\perp)$. $\Lambda$ and $\Lambda^\prime$ are the helicities of the initial-state baryon and the final-state baryon, respectively. $\lambda$ and $\lambda^\prime$ are the helicities of the initial-state struck quark and the final-state struck quark, respectively. One can decompose the light-front helicity amplitudes, Eq.~(\ref{TMDsoverlap}), into the TMDs as~\cite{Bacchetta:1999kz}
\begin{widetext}
    \begin{equation}
        \Phi_{\Lambda^{\prime} \lambda^{\prime}, \Lambda \lambda;q}=\left(\begin{array}{cccc}
        \frac{1}{2}\left(f_{1}^{q}+g_{1 L}^{q}\right) & -\frac{k_{R}}{2 M}\left(i h_{1}^{\perp q}-h_{1 L}^{\perp q}\right) & \frac{k_{L}}{2 M}\left(i f_{1 T}^{\perp q}+g_{1 T}^{q}\right) & h_{1}^{q} \\
        \frac{k_{L}}{2 M}\left(i h_{1}^{\perp q}+h_{1 L}^{\perp q}\right) & \frac{1}{2}\left(f_{1}^{q}-g_{1 L}^{q}\right) & \frac{k_{L}^{2}}{2 M^{2}} h_{1 T}^{\perp q} & \frac{k_{L}}{2 M}\left(i f_{1 T}^{\perp q}-g_{1 T}^{q}\right) \\
        -\frac{k_{R}}{2 M}\left(i f_{1 T}^{\perp q}-g_{1 T}^{q}\right) & \frac{k_{R}^{2}}{2 M^{2}} h_{1 T}^{\perp q} & \frac{1}{2}\left(f_{1}^{q}-g_{1 L}^{q}\right) & -\frac{k_{R}}{2 M}\left(i h_{1}^{\perp q}+h_{1 L}^{\perp q}\right) \\
        h_{1}^{q} & -\frac{k_{R}}{2 M}\left(i f_{1 T}^{\perp q}+g_{1 T}^{q}\right) & \frac{k_{L}}{2 M}\left(i h_{1}^{\perp q}-h_{1 L}^{\perp q}\right) & \frac{1}{2}\left(f_{1}^{q}+g_{1 L}^{q}\right)
        \end{array}\right),
    \end{equation}
\end{widetext}
where $k_{R,L}=k^1_\perp\pm ik^2_\perp$. The row indices are the final-state light-front helicities of the baryon and the struck quark $(\Lambda^\prime,\lambda^\prime)=$ $(+,+)$, $(+,-)$, $(-,+)$, $(-,-)$, while the column indices are the initial-state light-front helicities $(\Lambda,\lambda)$=$(+,+)$, $(+,-)$, $(-,+)$, $(-,-)$.

In this work, we focus on the internal structure of $\Lambda$ and $\Lambda_c$ baryons. We treat the baryons as composite particles composed only of the valence quarks. We expand the baryon bound state $|P,\Lambda\rangle$ in Fock space truncated to the leading Fock Sector in Eq.~(\ref{BoundState0}), where the index $i$ refers to the flavor index ($i=u,d,s$, for $\Lambda$; $i=u,d,c$, for $\Lambda_c$). 

We employ the hermiticity properties of TMDs and ignore the gauge links, $\Phi^*_{\Lambda^\prime \lambda^\prime,\Lambda \lambda}=\Phi_{\Lambda \lambda,\Lambda^\prime \lambda^\prime}$. Further, the helicities flip symmetry of the LFWFs in BLFQ~\cite{Xu:2021wwj} is
\begin{equation}
    \Psi^\Lambda_{\lambda_1,\lambda_2,\lambda_3}=(-)^{\frac{\Lambda-\lambda_1-\lambda_2-\lambda_3}{2}+1}\Psi^{-\Lambda*}_{-\lambda_1,-\lambda_2,-\lambda_3}.
\end{equation}
Under the approximation that the gauge link is the identity operator, the TMDs are obtained by the following overlaps of the three-quark LFWFs
\begin{align}
    f_1=\sum_{\lambda_1\lambda_2 \lambda_3}\int[D]\left[|\Psi^{+}_{\lambda_1\lambda_2\lambda_3}|^2+|\Psi^{-}_{\lambda_1\lambda_2\lambda_3}|^2\right]\label{TMD1},\\
    g_{1L}=\sum_{\lambda_2 \lambda_3}\int[D]\left[|\Psi^{+}_{+\lambda_2\lambda_3}|^2-|\Psi^{+}_{-\lambda_2\lambda_3}|^2\right]\label{TMD2},\\
    f^{\perp}_{1T}=0,\label{TMD3}\\
    g_{1T}=\frac{2M}{|\vec k_\perp|^2}\sum_{\lambda_2 \lambda_3}\int[D]\left[\Re \left(k_R\Psi^{+*}_{+\lambda_2\lambda_3}\Psi^{-}_{+\lambda_2\lambda_3}\right)\right],\\
    h_1=\sum_{\lambda_2 \lambda_3}\int[D]\Re\left[\Psi^{+*}_{+\lambda_2\lambda_3}\Psi^{-}_{-\lambda_2\lambda_3}\right],\\
    h^{\perp}_{1T}=\frac{2M^2}{|\vec k_\perp|^4}\sum_{\lambda_2 \lambda_3}\int[D]\Re \left[k^2_L\Psi^{+*}_{-\lambda_2\lambda_3}\Psi^{-}_{+\lambda_2\lambda_3}\right],\\
    h_1^{\perp}=0,\\
    h^{\perp}_{1L}=\frac{2M}{|\vec k_\perp|^2}\sum_{\lambda_2 \lambda_3}\int[D]\left[\Re \left(k_L\Psi^{+*}_{+\lambda_2\lambda_3}\Psi^{+}_{-\lambda_2\lambda_3}\right)\right],
        \label{TMD8}
\end{align}
where $\displaystyle\int[D]\equiv\displaystyle\int\frac{\mathrm{d}x_2\mathrm{d}^2\vec k_{\perp2}}{2(2\pi)^3(2\pi)^3}$. We omit the variables $(x,k^2_\perp)$ of the TMDs and $(\{\tilde{k}_i\})$ of the LFWFs, where $\tilde{k}_3=(1-x-x_2,-\vec k_\perp-\vec k_{\perp2})$ is owing to the $\delta$ functions in Eq.~(\ref{TMDsoverlap}) and the conservation of momentum. It is worth mentioning that the six T-even TMDs under the BLFQ framework are mutually independent~\cite{Hu:2022ctr}, while in other models they are related~\cite{Avakian:2010br,Pasquini:2012jm,Lorce:2011zta}.

\subsection{Probabilistic interpretations\label{Probabilistic_interpretations}}
From the overlap forms of $f_1$ in Eq.~(\ref{TMD1}) and $g_{1L}$ in Eq.~(\ref{TMD2}), we know their probabilistic interpretations. The $f_1$ describes the distribution of unpolarized quarks with the given momentum $\tilde{k}$ in an unpolarized hadron; the $g_{1L}$ describes 
the difference between the number densities of quarks with the positive and negative helicities in a longitudinally polarized hadron.
However, the probabilistic interpretations of the other TMDs are hidden in their overlap forms, especially $f^\perp_{1T}$ and $h_1^\perp$. In Refs.~\cite{Tangerman:1994eh,Barone:2001sp}, the authors summarized the probabilistic interpretations of the twist-2 TMDs by analyzing the bilocal operator structure and the parameterization of the TMD correlators. In this section, we review the probabilistic interpretations of $f_1(x,k_\perp^2)$, $g_{1L}(x,k_\perp^2)$, and $h_1(x,k_\perp^2)$.

Employing the helicity projection operator $\hat{O}_{R/L}\equiv\frac{1}{4}(1\pm\gamma^5)$ and the transverse polarization projection operator $\hat{O}_{\uparrow/\downarrow}\equiv\frac{1}{4}(1\pm\gamma^1\gamma^5)$, the Dirac matrices in the leading-twist TMD correlators project the bilocal quark operator into
\begin{align}
    \bar\psi(0) \gamma^+\psi(z)&=2\psi^\dagger_+(0)\psi_+(z),\label{psigamma+psi}\\
    \bar\psi(0) \gamma^+\gamma^5\psi(z)&=2(\psi^\dagger_{+,R}(0)\psi_{+,R}(z)-\psi^\dagger_{+,L}(0)\psi_{+,L}(z)),\\
    \bar\psi(0) i\sigma^{1+}\gamma^5\psi(z)&=2(\psi^\dagger_{+,\uparrow}(0)\psi_{+,\uparrow}(z)-\psi^\dagger_{+,\downarrow}(0)\psi_{+,\downarrow}(z))\label{psisigmaj+psi},
\end{align}  
where $\psi_+$ represents the `good' field. $\psi_R$ and $\psi_L$ represent the states of the quark field with positive and negative helicity, respectively. $\psi_\uparrow$ and $\psi_\downarrow$ denote the states of the quark field with transverse polarisation $\uparrow$ and transverse polarisation $\downarrow$, respectively.

Substituting Eqs.~(\ref{psigamma+psi}-\ref{psisigmaj+psi}) into the correlators of the left hand side of Eqs.~(\ref{TMDs1}-\ref{TMDs3}), inserting a complete set of on-shell intermediate states $\{|n\rangle\}$, and employing the translation operator, we have 
\begin{align}
    \Phi^{[\gamma^+]}&=\sum_n \delta^3({P}-{k}-{p}_n)|\langle P,S|\psi_+(0)|n\rangle|^2\label{Phigammaplus},\\
    \Phi^{[\gamma^+\gamma^5]}&=\sum_n \delta^3({P}-{k}-{p}_n)\{|\langle P,S|\psi_{+,R}(0)|n\rangle|^2\nonumber\\
    &-|\langle P,S|\psi_{+,L}(0)|n\rangle|^2\},\label{Phigammaplusgamma5}\\
    \Phi^{[i\sigma^{1+}\gamma^5]}&=\sum_n \delta^3({P}-{k}-{p}_n)\{|\langle P,S|\psi_{+,\uparrow}(0)|n\rangle|^2\nonumber\\
    &-|\langle P,S|\psi_{+,\downarrow}(0)|n\rangle|^2\}\label{Phisigma1plus},
\end{align}
where $\delta^3({P}-{k}-{p}_n)\equiv\delta(P^+-xP^+-p_n^+)\delta^2(\vec P_\perp-\vec k_\perp-\vec p_{\perp n})$. $\sum_n$ represents summing over the phase space of all the intermediate states $|n\rangle$, including $|qq\rangle$, $|qqg\rangle$, $|qqq\bar q\rangle$, etc. In Eq.~(\ref{Phigammaplus}), we know that the TMD correlator $\Phi^{[\gamma^+]}$ means the probability density of finding an unpolarized quark with the longitudinal momentum fraction $x$ and the transverse momentum $k_\perp$ inside a hadron with the spin $S$ and the four-momentum $P$. In Eq.~(\ref{Phigammaplusgamma5}), the TMD correlator $\Phi^{[\gamma^+\gamma^5]}$ denotes the difference between the densities of quarks with positive helicity and with negative helicity in a hadron. In Eq.~(\ref{Phisigma1plus}), the TMD correlator $\Phi^{[i\sigma^{j+}\gamma^5]}$ represents the difference between the densities of quarks with different transverse polarizations in a hadron.

Through the analysis of the structure of the above bilocal operators, we know the meaning of the TMD correlators. 
In Ref.~\cite{Barone:2001sp}, the authors express those leading-twist TMD correlators in the entries of the spin density matrix of quarks in the baryon. Here, we define 
\begin{equation}
    \mathcal{P}_{q,s/H,S}(x,\vec k_\perp)=\sum_n \delta^3({P}-{k}-{p}_n)|\langle P,S|\psi_{+,s}(0)|n\rangle|^2,
\end{equation}
which represents the probability density of finding a $s$-polarized quark with $(x,\vec k_\perp)$ in a hadron with polarization $S$. Then the TMD correlators in Eqs.~(\ref{TMDs1}-\ref{TMDs3}) are
\begin{align}
    \Phi^{[\gamma^+]}&=\mathcal{P}_{q/H,S}(x,\vec k_\perp)\nonumber\\
    &=f_1(x,k^2_\perp)-\frac{\epsilon_{\perp}^{i j} k_\perp^{i} S_\perp^{j}}{M} f_{1 T}^{\perp}(x,k^2_\perp),\\
    \Phi^{\left[\gamma^{+} \gamma^{5}\right]}&=\mathcal{P}_{q,+/H,S}(x,\vec k_\perp)-\mathcal{P}_{q,-/H,S}(x,\vec k_\perp)\nonumber\\
    &=S^{3} g_{1 L}(x,k^2_\perp)+\frac{\vec k_{\perp} \cdot \vec S_\perp}{M} g_{1 T}(x,k^2_\perp),\\
    \Phi^{\left[i \sigma^{j+} \gamma^{5}\right]}&=\mathcal{P}_{q,\uparrow/H,S}(x,\vec k_\perp)-\mathcal{P}_{q,\downarrow/H,S}(x,\vec k_\perp)\nonumber\\
    &=S_\perp^{j} h_{1}(x,k^2_\perp)+S^{3} \frac{k_\perp^{j}}{M} h_{1 L}^{\perp}(x,k^2_\perp)\nonumber\\
    &+S_\perp^{i} \frac{2 k_\perp^{i} k_\perp^{j}-\left(\vec k_{\perp}\right)^{2} \delta^{i j}}{2 M^{2}} h_{1 T}^{\perp}(x,k^2_\perp)\nonumber\\
    &+\frac{\epsilon_{\perp}^{j i} k_\perp^{i}}{M} h_{1}^{\perp}(x,k^2_\perp),
\end{align}
where $\mathcal{P}_{q/H,S}$ represents the probability density of an unpolarized quark.

In the infinite-momentum frame, we define the azimuth angles of the transverse momentum $\phi_k$, the hadron spin $\phi_S$, and the quark spin $\phi_s$ in the plane orthogonal to the direction of hadron motion, respectively. After integrating the TMD correlators over $\phi_k$, the other TMDs without collinear interpretations disappear, 
\begin{align}
    \mathcal{P}_{q/H}(x,\vec k_\perp)&=f_1(x, k_\perp^2),\\
    \mathcal{P}_{q,+/H,+}(x,\vec k_\perp)-\mathcal{P}_{q,-/H,+}(x,\vec k_\perp)&=g_{1L}(x, k_\perp^2),\\
    \mathcal{P}_{q,\uparrow/H,\uparrow}(x,\vec k_\perp)-\mathcal{P}_{q,\downarrow/H,\uparrow}(x,\vec k_\perp)&=\cos(\phi_S-\phi_s)h_1(x, k_\perp^2),
\end{align}
where the labels `$+$/$-$' and `$\uparrow$/$\downarrow$' denote longitudinal and transverse polarization, respectively.

The above derivation is model independent. It reveals the probabilistic meaning of TMDs from the perspective of field operators. $f_1$ describes the distribution of unpolarized quarks; $g_{1L}$ ($h_{1}$) describes the difference of the distribution of longitudinally (transversely) polarized quarks in a longitudinally (transversely) polarized baryon.

\subsection{Inequality relations}
In Sec.~\ref{Overlap_representation}, we derived the LFWF overlap forms of the TMDs. After flipping the helicity of the LFWF in the unpolarized TMD in Eq.~(\ref{TMD1}),
\begin{equation}
    f_1=\sum_{\lambda_1\lambda_2 \lambda_3}\int\frac{\mathrm{d}x_2\mathrm{d}^2\vec k_{\perp 2}}{2(2\pi)^3(2\pi)^3}\left[|\Psi^{+}_{+\lambda_2\lambda_3}|^2+|\Psi^{+}_{-\lambda_2\lambda_3}|^2\right]\label{TMD1prime},
\end{equation}
we find the bound relation between $f_1(x,k_\perp^2)$ and $g_{1L}(x,k_\perp^2)$ from their overlap representations in Eqs.~(\ref{TMD2},~\ref{TMD1prime}),
\begin{equation}
    |g_{1L}(x,k_\perp^2)|\leq f_1(x,k_\perp^2)\label{ineq1}.
\end{equation}
In addition, the T-even twist-2 TMDs have other bounds~\cite{Soffer:1994ww,Bacchetta:1999kz}, 
\begin{align}
    0&\leq f_1(x,k_\perp^2),\\
    |h_1(x,k_\perp^2)|&\leq f_1(x,k_\perp^2),\\
    |h_1(x,k_\perp^2)|&\leq \frac{1}{2}|f_1(x,k_\perp^2)+g_{1L}(x,k_\perp^2)|,\\
    \frac{k^2_\perp}{2M^2}|h_{1T}^\perp(x,k_\perp^2)|&\leq \frac{1}{2}|f_1(x,k_\perp^2)-g_{1L}(x,k_\perp^2)|\label{ineq5}.
\end{align}
All the relations listed above are independent of any model. We test our results for consistency with those relations.

\section{NUMERICAL RESULTS\label{Sec4}}

According to Eqs.~(\ref{TMD1}-\ref{TMD8}), our results for the valence quark TMDs of $\Lambda$ and $\Lambda_c$ baryons are obtained from the overlaps of the three-quark LFWFs. In total, we have six model parameters: the light and heavy quark mass in the kinetic energy, $(m_{q/k})$, the light and heavy quark mass in the OGE interaction, $(m_{q/g})$, the strength of confining potential, $(\kappa)$, and the coupling constant, $(\alpha_s)$, in the OGE interaction~\cite{Mondal:2019jdg,Xu:2021wwj}, and we select three computational parameters: the HO scale parameter $b=0.6\;\text{GeV}$ and the truncation parameters $N_\text{max}=8$, $K=16.5$. With the model parameters mentioned in Table~\ref{table-parameters} in Ref.~\cite{Peng:2022lte}, we identify the ground state as the $\Lambda$ and $\Lambda_c$ baryon and get the TMDs from the LFWFs of the $\Lambda$ and $\Lambda_c$ baryon with masses $M_\Lambda=1.116$ GeV and $M_{\Lambda_c}=2.287$ GeV, respectively.
\begin{table}[!h]
	\caption{List of the model parameters with the truncation $\{N_{\text{max}},K\}=\{8,16.5\}$ for $\Lambda$, $\Lambda_c$~\cite{Peng:2022lte} and the proton~\cite{Xu:2021wwj}. All are in units of GeV except $\alpha_s$.}
	\centering 
		\begin{tabular}{cccccc}
			\hline \hline
			&  $\alpha_s$& $m_{q/k}/m_{q/g}$ &$m_{s/k}/m_{s/g}$& $m_{c/k}/m_{c/g}$ & $\kappa$ \\ 
			\hline 		
			$\Lambda$&  1.06&  0.30/0.20& 0.39/0.29 & - &  0.337 \\ 		
			$\Lambda_c$&  0.57&  0.30/0.20& - & 1.58/1.48&   0.337 \\ 
            proton& 1.10 &0.30/0.20  & - & - & 0.337 \\ 
			\hline \hline 	
		\end{tabular}
	\label{table-parameters} 
\end{table}
\subsection{The TMDs of valence quarks}

\begin{figure*}
    \includegraphics[width=0.98\textwidth]{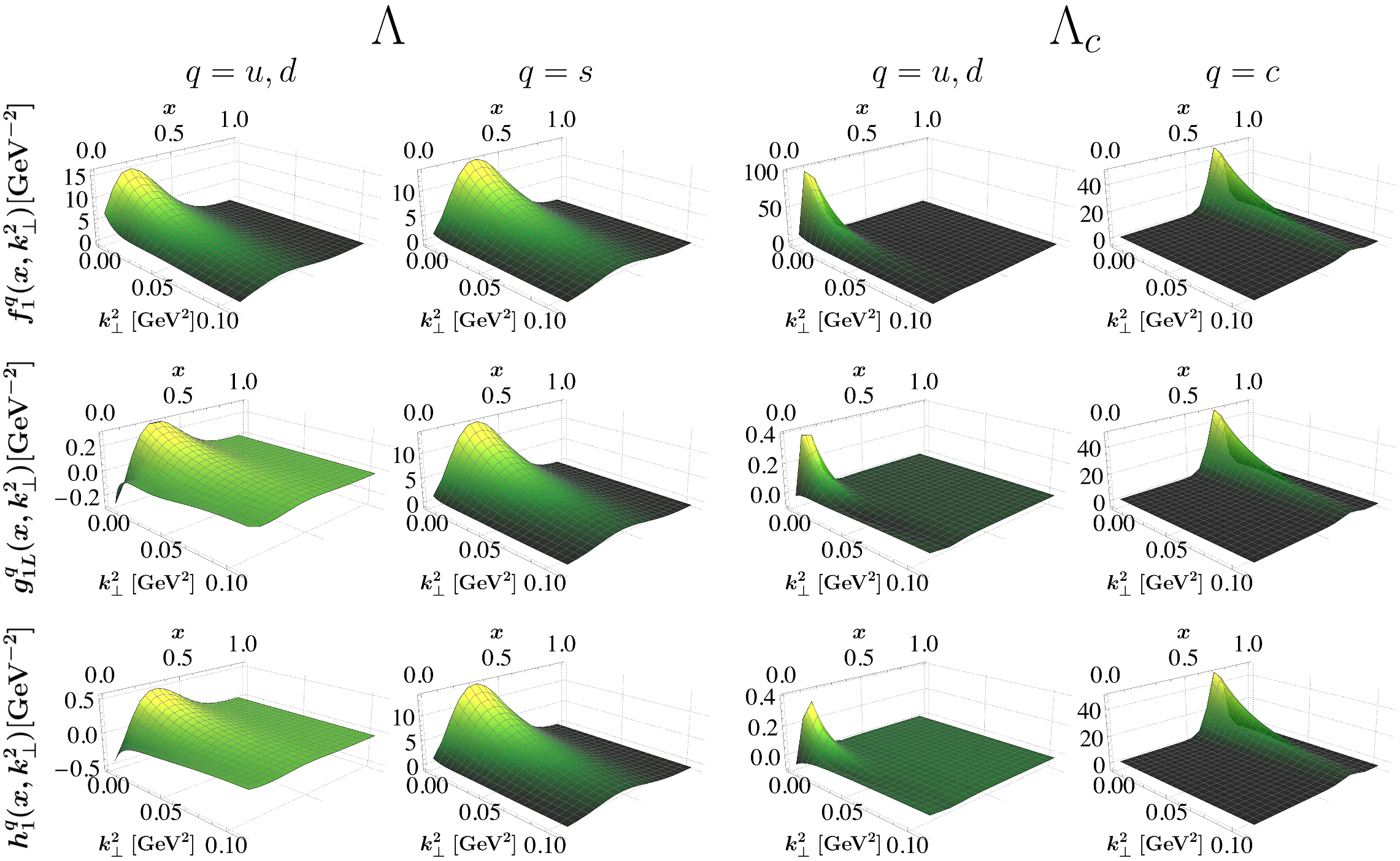} 
    \caption{Three-dimensional plots for TMDs of the light quarks and the $s$ and $c$ quark inside $\Lambda$ (two columns on the left) and $\Lambda_c$ baryons (two columns on the right) ignoring the gauge link. The images from the first row to the third row are the unpolarized TMD $f_1^q$, the helicity TMD $g^q_{1L}$, and the transversity TMD $h^q_1$, respectively. The BLFQ computations are carried out at $N_{\text{max}}=8$ and $K=16.5$. }
        \label{TMD_lambdalambdac}
\end{figure*}
    
Figure~\ref{TMD_lambdalambdac} shows our model results for the T-even TMDs without evolution effects or gauge links for the valence quarks inside the $\Lambda$ and $\Lambda_c$ baryons in the BLFQ framework. Our results satisfy the inequality relations in Eqs.~(\ref{ineq1}-\ref{ineq5}). We only show results for the $u$ quark instead of both the light quarks. The reason is that the light quark's TMDs are nearly identical, since they have the same mass and they have the same structure in the flavor-spin symmetry analysis (see Sec.~\ref{SpinStructure}). It is worth noting that in the spin analysis of the quark model, the spatial wave function contains only the S-wave in Sec.~\ref{SpinStructure}. However, our results contain not only the S-wave contributions but also the combined contributions from P-waves and D-waves, but for the $\Lambda$ ($\Lambda_c$) baryon, the former contributes $74\%$ ($96\%$) significantly larger than the latter two combined $26\%$ ($4\%$). Therefore, the conclusions of Sec.~\ref{SpinStructure} apply well to our results. 

\begin{figure*}
    \includegraphics[width=0.7\textwidth]{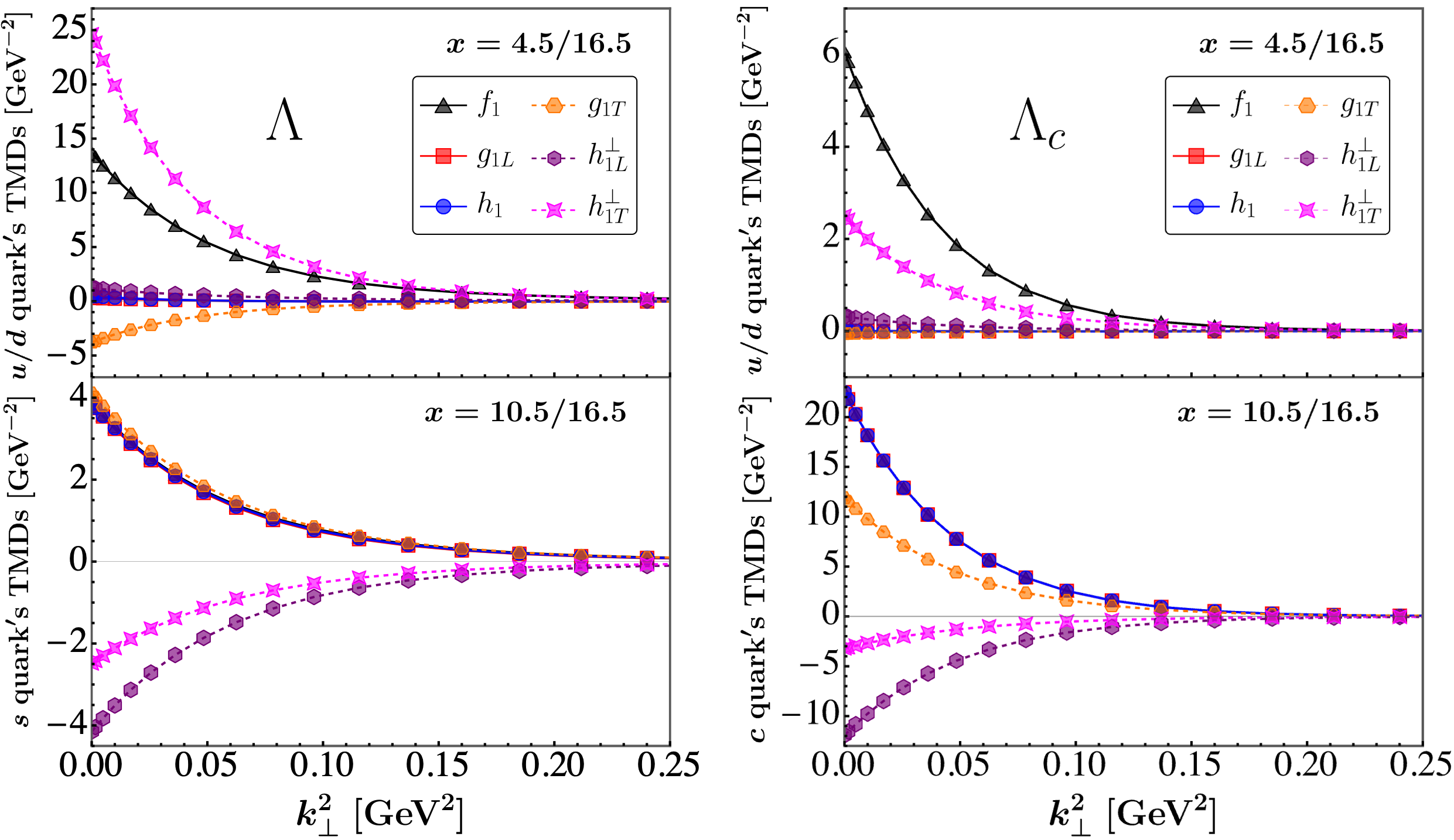}  
    \caption{Two-dimensional plots for the $k^2_\perp$-dependence of TMDs at fixed $x$ for the valence quarks inside $\Lambda$ and $\Lambda_c$ baryons. The BLFQ computations are carried out at $N_{\text{max}}=8$ and $K=16.5$. }
    \label{kT-TMD_lambda}
\end{figure*}

\begin{figure*}
    \includegraphics[width=0.7\textwidth]{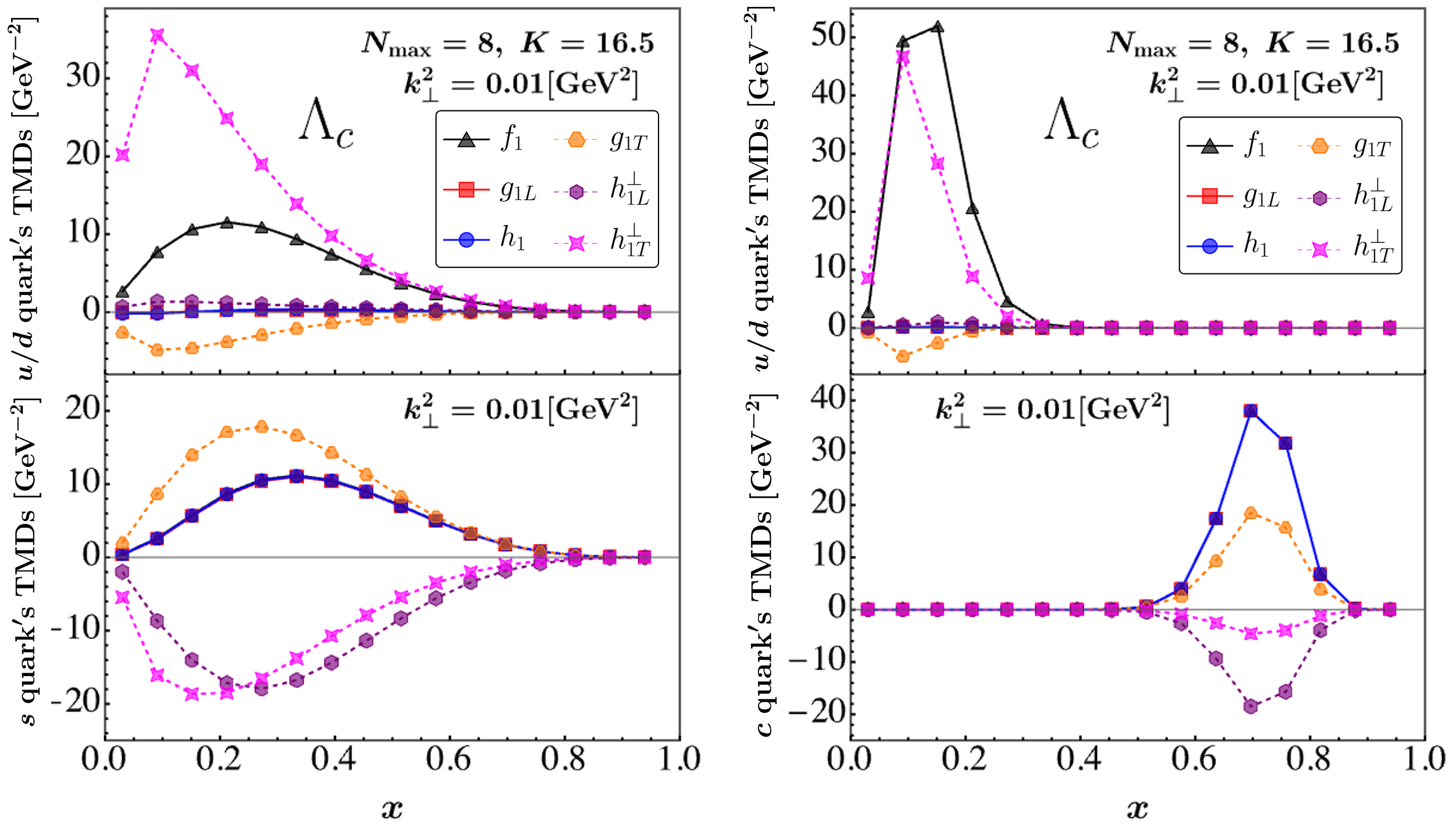} 
    \caption{Two-dimensional plots for the $x$-dependence of TMDs at fixed $k^2_\perp$ for the valence quarks inside $\Lambda$ and $\Lambda_c$ baryons. The BLFQ computations are carried out at $N_{\text{max}}=8$ and $K=16.5$.}
    \label{x-TMD_lambda}
\end{figure*}

For clarity, the 2-D sections in the transverse momentum $k_\perp$-direction and the longitudinal momentum fraction $x$-direction of these TMDs are shown in Fig.~\ref{kT-TMD_lambda} and Fig.~\ref{x-TMD_lambda} for the $\Lambda$ baryon and the $\Lambda_c$ baryon, respectively.

Figure~\ref{kT-TMD_lambda} shows the quark TMDs for the $\Lambda$ and $\Lambda_c$ baryons as functions of $k^2_\perp$ at fixed $x=4.5/16.5$ for light quarks, and $x=10.5/16.5$ for $s$ and $c$ quarks, respectively. For the $s$ and $c$ quarks, we select a different fixed $x=10.5/16.5$, since the $c$ quark TMDs for $\Lambda_c$ concentrate at a larger $x$ region and this choice makes them visually clearer. All the TMDs for the $\Lambda$ and $\Lambda_c$ baryons at any fixed $x$ have their peaks at the transverse momentum $ k^2_\perp=0$, consistent with the dominance of the S waves.

Figure~\ref{x-TMD_lambda} shows the quark TMDs for the $\Lambda$ and $\Lambda_c$ baryons as functions of $x$ at fixed $k_\perp^2=0.01\;\rm{GeV^2}$. The plots reveal a peak or a valley structure near $x\approx0.2\;(0.1)$ for the light quarks, and larger $x\approx0.3\;(0.7)$ for the $s$ ($c$) quark inside the $\Lambda$ ($\Lambda_c$) baryon, respectively. The reason is that the heavy quarks carry more longitudinal momentum fraction in the bound system. We can find that the helicity and the transversity TMDs of light quarks are almost zero, while those of $s$ and $c$ quarks are comparable to the unpolarized TMD $f_1$. This is understandable since, in the $\Lambda$ and $\Lambda_c$ baryons dominated by S waves, the light quarks are unpolarized with similar densities of positive or negative helicity. In addition, the valence quark inside the $\Lambda_c$ baryon has a narrow longitudinal distribution compared with the $\Lambda$ baryon, since the mass of the $c$ quark in the $\Lambda_c$ baryon is much heavier than that of the light quarks. Also due to its large masses, the heavy quark is more concentrated in the large-$x$ region than the light quarks.

\subsection{The spin-density of valence quarks}

Twist-2 TMDs have a probabilistic interpretation, as discussed above, but the probability density of different polarizations is mixed~\cite{Tangerman:1994eh,Barone:2001sp}. To understand the full 3-D dynamics of partons in the composite system, we discuss the spin densities of valence quarks in the transverse momentum plane~\cite{Pasquini:2012jm},
\begin{align}
    \rho&\left(\vec k_\perp,\pmb s,\pmb S\right)=\notag\\
    \frac{1}{2} &\left[f_{1}+{S_{\perp}^{i} \epsilon^{i j} k^{j}_\perp \frac{1}{M} f_{1 T}^{\perp}}+\lambda \Lambda g_{1 L}+\lambda S_{\perp}^{i} k^{i}_\perp \frac{1}{M} g_{1 T}\right.\notag\\
    &+{s_{\perp}^{i} \epsilon^{i j} k^{j}_\perp \frac{1}{M} h_{1}^{\perp}}+\Lambda s_{\perp}^{i} k^{i}_\perp \frac{1}{M} h_{1 L}^{\perp}+s_{\perp}^{i} S_{\perp}^{i} h_{1}\notag \\
    &\left.+s_{\perp}^{i}\left(2 k^{i}_\perp k^{j}_\perp-\vec{k}_{\perp}^{2} \delta^{i j}\right) S_{\perp}^{j} \frac{1}{2 M^{2}} h_{1 T}^{\perp}\right],
    \label{SpinDensity}
\end{align}
where $M$ represents the baryon mass. $\pmb s=(\lambda,\vec s_\perp)$ and $\pmb S=(\Lambda,\vec S_\perp)$ are the spin of the struck quark and the baryon, respectively. For a generic TMD $j$, we introduce the $x$-integrated function defined as~\cite{Pasquini:2012jm}
\begin{equation}
    j(k^2_\perp)=\int \mathrm{d}x j(x,k^2_\perp).
\end{equation}

From Eq.~(\ref{SpinDensity}) and omitting the T-odd TMDs $f_{1 T}^{\perp}$ and $h_{1}^{\perp}$, we can form the following six densities of quarks with different polarization of the struck quark and the hadron ignoring the gauge links. For an unpolarized baryon, we define the unpolarized spin-density
\begin{equation}
    \rho\left(k_{x}, k_{y}\right)=f_{1}^{q}\label{rho_UU},
\end{equation}
as the probability density of finding an unpolarized quark with a given $\vec{k}_{\perp}$ in an unpolarized baryon.

The helicity density
\begin{equation}
    \rho^{+/+}\left(k_{x}, k_{y}\right)=\frac{1}{2}\left(f_{1}^{q}+g_{1 L}^{q}\right)\label{rho_LL}
\end{equation}
represents the probability density of finding a longitudinally polarized quark with $\vec{k}_{\perp}$ in a longitudinally polarized baryon.

The transversity density 
\begin{equation}
    \begin{aligned}
    \rho^{\uparrow / \uparrow}\left(k_{x}, k_{y}\right)=\frac{1}{2} &\left(f_{1}^q+h_{1}^q+\frac{k^2_y-k^2_x}{2M^2} h_{1 T}^{\perp q}\right)
    \label{rho_TT}
    \end{aligned}
\end{equation}
describes the probability density of finding a transversely polarized quark in the baryon with the same transverse polarization.

In the baryon with transverse polarization along the $y$-axis, the Worm-gear density
\begin{equation}
    \rho^{+ / \uparrow}\left(k_{x}, k_{y}\right)=\frac{1}{2}\left(f_{1}^{q}+\frac{k_{y}}{M} g_{1 T}^{q}\right)\label{rho_LT}
\end{equation}
refers to the probability density of finding a longitudinally polarized quark.

The Kotzinian-Mulders density
\begin{equation}
    \rho^{\uparrow / +}\left(k_{x}, k_{y}\right)=\frac{1}{2}\left(f_{1}^{q}+\frac{k_y}{M}h_{1L}^{\perp q}\right)\label{rho_TL}
\end{equation}
represents the probability density of finding a quark with transverse polarization along the $y$-axis in a longitudinally polarized baryon.

Finally, the Pretzelosity density
\begin{equation}
    \rho^{\uparrow / \rightarrow}\left( k_{x}, k_{y}\right)=\frac{1}{2}\left(f_{1}^{q}+\frac{k_x k_y}{M^2}h^{\perp q}_{1T}\right)\label{rho_UR}
\end{equation}
means the probability density of finding a quark with different transverse polarization to the baryon.

The first and second superscripts of the above spin-densities indicate the respective polarization of the struck quark and the baryon. The label `+' refers to longitudinal polarization. The label `$\uparrow$/$\rightarrow$' means transverse polarization along the $y/x$-axis.

\begin{figure*}
    \includegraphics[width=0.94\textwidth]{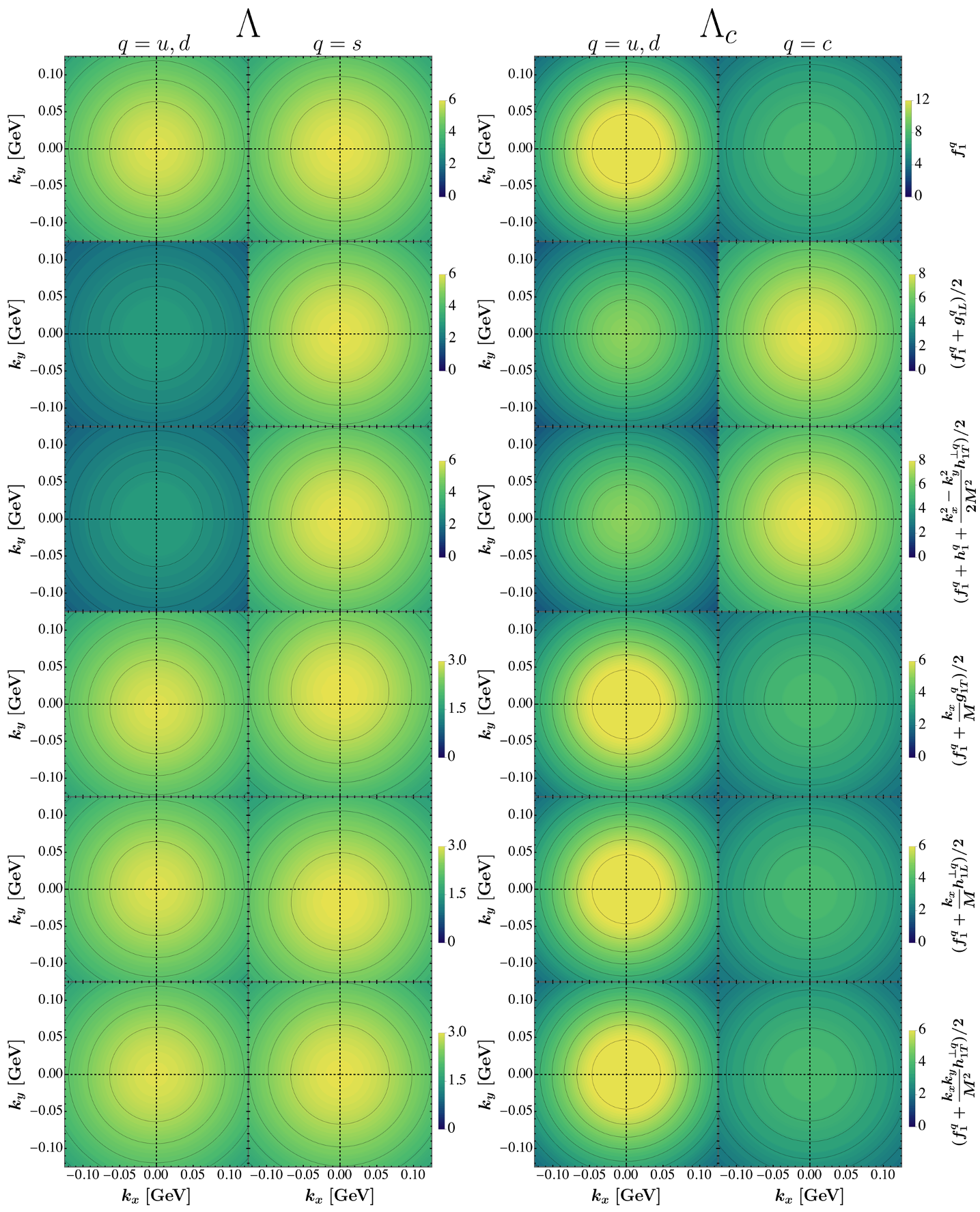}
    \caption{The density plots of valence quarks inside the $\Lambda$ and $\Lambda_c$ baryons in the transverse-momentum plane with different polarizations. The plots in the first row are the unpolarized density, Eq.~(\ref{rho_UU}), as the probability density of finding unpolarized quarks at a given $\vec k_\perp$ in the unpolarized baryon. The plots in the second row are the helicity density, Eq.~(\ref{rho_LL}), as the probability density of finding longitudinally polarized quarks in the baryon with the same longitudinal polarization. The plots in the third row are the transverse density, Eq.~(\ref{rho_TT}), as the probability density of finding transversely polarized quarks in the baryon with the same transverse polarization in $y$-axis. The plots in the fourth row are the Worm-gear density, Eq.~(\ref{rho_LT}), as the probability density of finding longitudinally polarized quarks in the $y$-axis transversely polarized baryon. The plots in the fifth row are the probability density, Eq.~(\ref{rho_TL}), of finding $y$-axis transversely polarized quarks in the longitudinally polarized baryon. All the densities are in units of $\mathrm{GeV^{-2}}$. The BLFQ computations are carried out at $N_\text{max}=8$ and $K=16.5$.}
    \label{density_lambda_0.1}
\end{figure*}
    
In Fig.~\ref{density_lambda_0.1}, we show the spin densities of the valence quarks inside the $\Lambda$ and $\Lambda_c$ baryons in the transverse-momentum plane in the BLFQ framework. The contour plots in the first row of Fig.~\ref{density_lambda_0.1} show the unpolarized spin density defined in Eq.~(\ref{rho_UU}). The unpolarized density is independent of the transverse azimuth angle. As expected, it is circularly symmetric around the direction of the baryon perpendicular to the paper's surface. Due to the same reason, the helicity spin density given in Eq.~(\ref{rho_LL}) in the second row also has a circular symmetry in the transverse-momentum plane. Whereas the transverse density in Eq.~(\ref{rho_TT}) involves a slight distortion term similar to quadrupole moment, thus $\rho^{\uparrow/\uparrow}$ for the valence quarks shows an elliptical structure whose major axis is the polarization axis, the $y$-axis, especially for the light quarks inside $\Lambda$. 

In the fourth row of Fig.~\ref{density_lambda_0.1}, the contour plots show the Worm-gear density defined in Eq.~(\ref{rho_LT}). For the $\Lambda$ baryon, it has a slight shift in the $y$-axis because the baryon is transversely polarized along the $k_y$ direction, and the longitudinal polarization of the quark does not affect the transverse azimuth distribution. In the fifth row of Fig.~\ref{density_lambda_0.1}, the $\rho^{\uparrow/+}$ density of the $s$ quark inside the $\Lambda$ baryon is asymmetric in the $y$-axis, but the $\rho^{\uparrow/+}$ densities of the light $u/d$ quarks almost have no shift. 
Because the Kotzinian-Mulders TMD $h^\perp_{1L}$ of the light quarks is smaller than $f^q_1$, the distortion is suppressed. 
In the bottom row, the $\rho^{\uparrow / \rightarrow}$ density also shows an elliptical structure whose major axis is in the line $k_y=k_x$ for the light quarks and $k_y=-k_x$ for the $s$ quark. 
The different asymmetry of the $\rho^{\uparrow / \rightarrow}$ density comes from the different signs of $h_{1T}^{\perp}$ for the light quarks and for the $s$ quark. 

For the $\Lambda_c$ baryon, the helicity density $\rho^{+/+}$ and the transversity density $\rho^{\rightarrow/\rightarrow}$ of the $c$ quark are almost twice as high as the unpolarized density. This is because the $c$ quark is scarcely antiparallel to the $\Lambda_c$ baryon, resulting in the antiparallel probabilities $\mathcal{P}_{c,-/\Lambda_c ,+}$ and $\mathcal{P}_{c,\downarrow/\Lambda_c,\uparrow}$ being small. The phenomenon also exists for the $s$ quark of the $\Lambda$ baryon. The other spin densities have no visible shifts. The reason is that all the TMDs for $\Lambda_c$ concentrate in small $|k_\perp|$ regions, and all the distortion terms contain transverse momentum factors, such as $k_y^2-k_x^2$ for the $\rho^{\rightarrow/\rightarrow}$. Thus, the distortion terms are suppressed by the small transverse momentum $k_\perp$. Owing to the spin structure of $\Lambda_c$, the TMDs of polarized light quarks are also very small.

\subsection{The comparison with protons}
Owing to the short lifetime of the $\Lambda$ and $\Lambda_c$ baryons, we qualitatively compare the TMDs with the stable and easy-to-detect proton in the same BLFQ framework. We adopt the model parameters for the proton mentioned in Table~\ref{table-parameters} in Refs.~\cite{Mondal:2019jdg,Mondal:2021wfq,Xu:2021wwj} and the same truncation $N_\text{max}=8$ and $K=16.5$.

\begin{figure*}
    \includegraphics[width=0.35\textwidth]{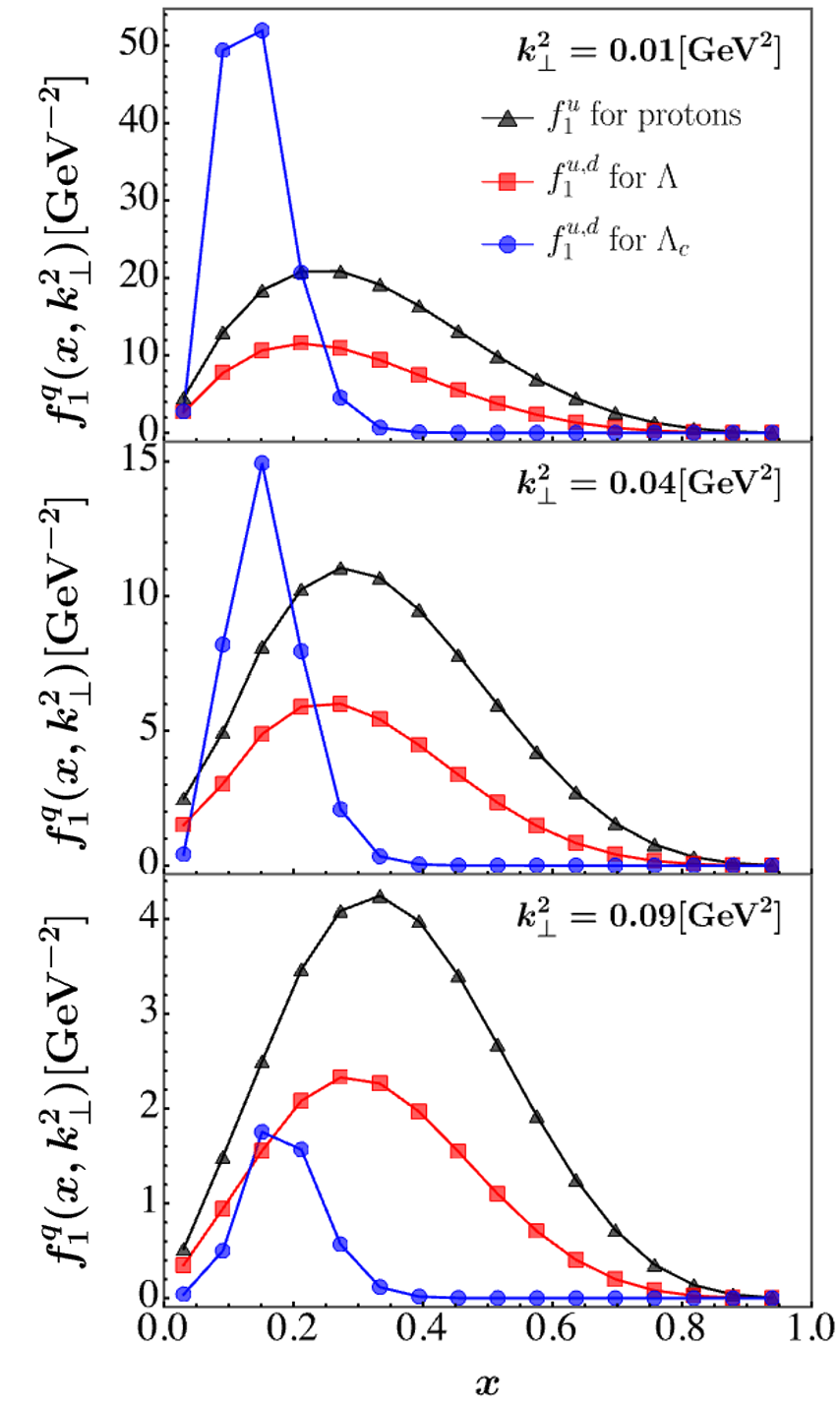}
    \includegraphics[width=0.35\textwidth]{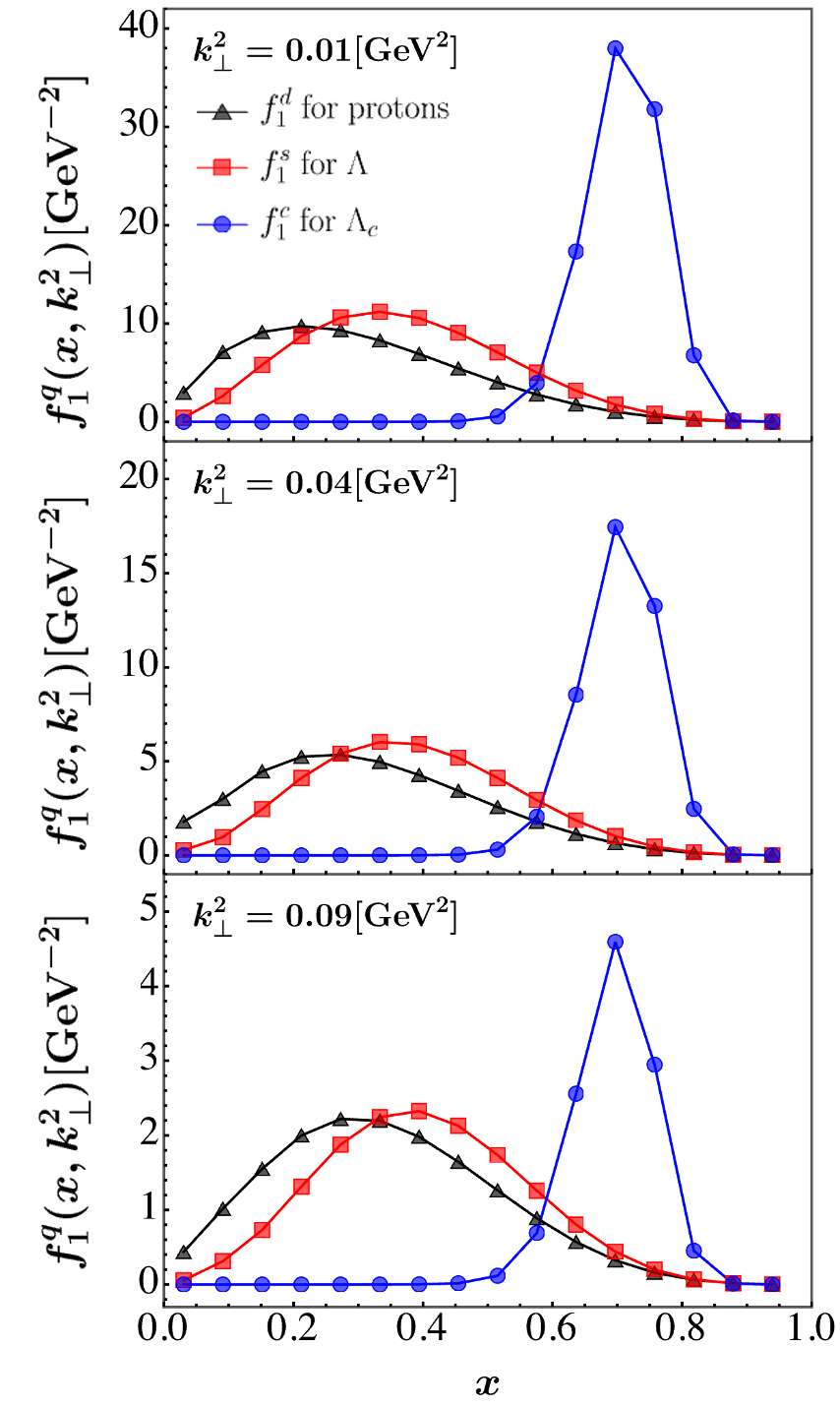}
    \caption{Two-dimensional plots for the $x$-dependence of the unpolarized TMD $f_1$ for light quarks and `heavy' quarks inside protons, $\Lambda$, and $\Lambda_c$. Light (heavy) quarks are compared in the left (right) column. For the proton, `light' refers to the $u$ quarks and `heavy' refers the $d$ quark. For $\Lambda$, the $s$ quark plays the role of the `heavy' quark.}
    \label{comparison_f1_x_direction}
\end{figure*}

Figure~\ref{comparison_f1_x_direction} compares the $x$-dependence of the unpolarized TMD $f_1$ of the light quarks and the $s$ and $c$ quarks at different $k^2_\perp$. It is clear that $f_1$ of the $u$ quarks inside the proton and $\Lambda$ have almost the same shape except for the magnitude, which differs by a factor of $2$. This is because we set the mass of the $u$ quarks inside the proton and $\Lambda$ to be the same and a proton has two valence $u$ quarks in the quark model. In addition, the proton and the $\Lambda$ baryon belong to the same baryon octet and have almost the same mass: the former is 938 MeV and the latter is 1115 MeV, which resulting in similar unpolarized structures. For $\Lambda_c$, the light quarks concentrate at smaller $x$ and distribute more narrowly than in the light baryons because relatively lighter quarks carry smaller longitudinal momentum fraction in a heavy baryon. In the right column in Fig.~\ref{comparison_f1_x_direction}, we find that the peak locations of $x$-dependence $f_1$ for the `heavy' quark are different. For the proton, the $d$ quark plays the role of the `heavy' quark. 
The heavier quarks have their peaks located at higher-$x$, revealing that they carry larger longitudinal momentum fraction.

While gradually increasing the transverse momentum $k^2_\perp$, the peak in amplitude of $x$-dependence of $f_1$ for the $\Lambda_c$ baryon drops rapidly compared to $\Lambda$ and the proton as illustrated in Fig.~\ref{comparison_f1_x_direction}. Considering the probabilistic interpretation of the quark unpolarized TMD $f_1$ in Sec.~\ref{Probabilistic_interpretations}, we define the mean squared transverse momentum of $f_1$ for quarks as 
\begin{equation}
    \langle k_\perp^2\rangle_{f_1}=\frac{\int\mathrm{d}x\int\mathrm{d}^2k_\perp k_\perp^2 f_1(x,k_\perp^2)}{\int\mathrm{d}x\int\mathrm{d}^2k_\perp  f_1(x,k_\perp^2)}.
\end{equation}
For $\Lambda$, $\Lambda_c$, and the proton, the mean $\langle k^2_\perp\rangle$ are summarized in Table~\ref{rms}. The $\langle k^2_\perp\rangle$ of the light quarks inside the $\Lambda_c$ baryon is almost twice that of $\Lambda$ and the proton. This is consistent with the physical intuition that the radius, $r\sim 1/k$, of the heavy system is smaller than that of the light system.
\begin{table}[ht]
	\caption{The mean square of the transverse momentum of the valence quarks in baryons. For $\Lambda$, the $s$ quark plays the role of the `heavy' quark. For the proton, `light' refers to the $u$ quarks and `heavy' refers the $d$ quark. All numbers are in units of $\mathrm{GeV^2}$. }
	\centering 
		\begin{tabular}{cccc}
			\hline \hline
            &  $\langle k^2_\perp\rangle_{\rm light} $&  $\langle k^2_\perp\rangle_{\rm heavy} $ &   \\ 
            \hline 		 		
			$\Lambda$&  0.067&   0.057& \\ 
			$\Lambda_c$& 0.116&   0.099&\\ 
            proton&  0.061&   0.071&   \\ 
			\hline \hline 	
		\end{tabular}
        \label{rms}
\end{table}

\begin{figure*}
    \includegraphics[width=0.375\textwidth]{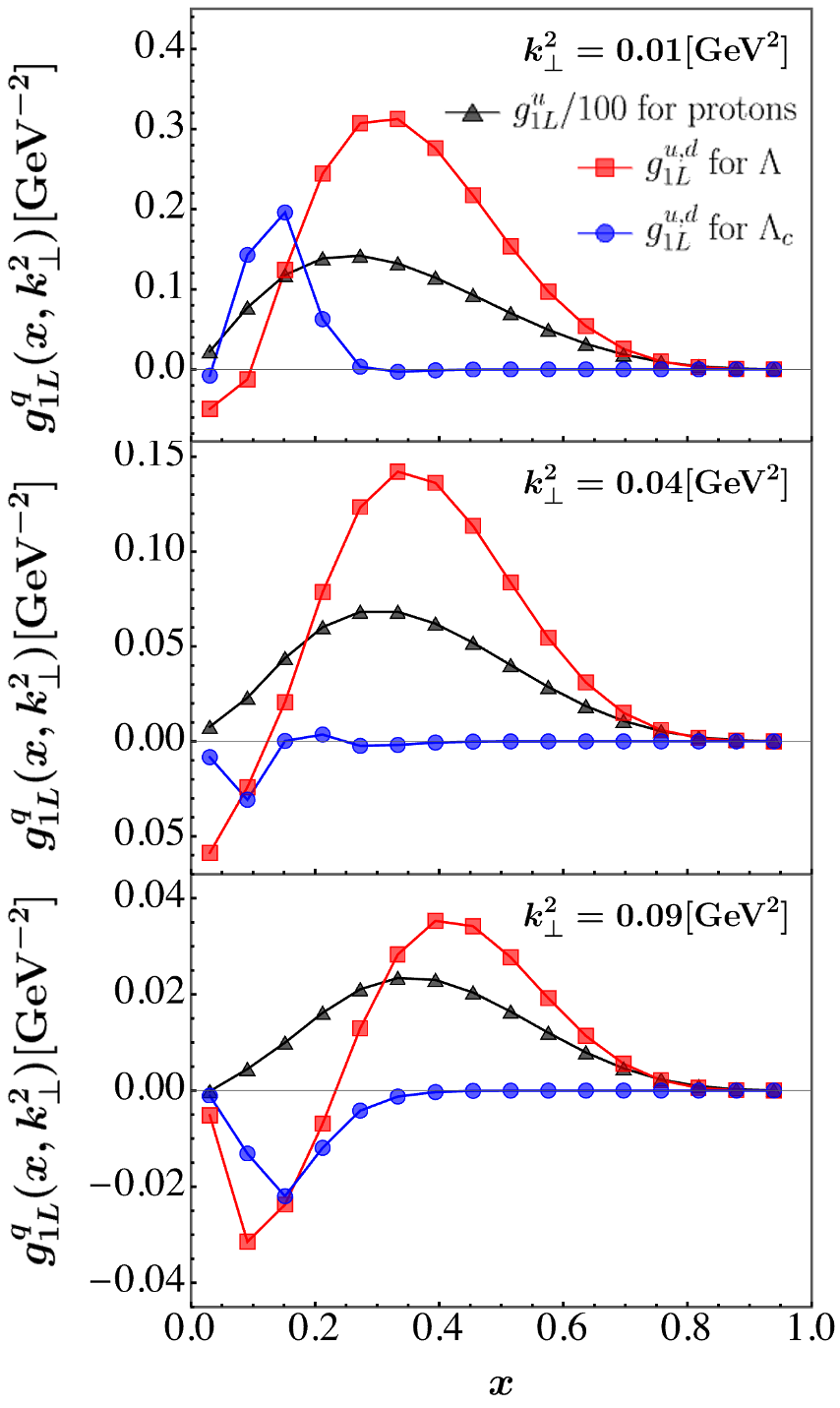}
    \includegraphics[width=0.375\textwidth]{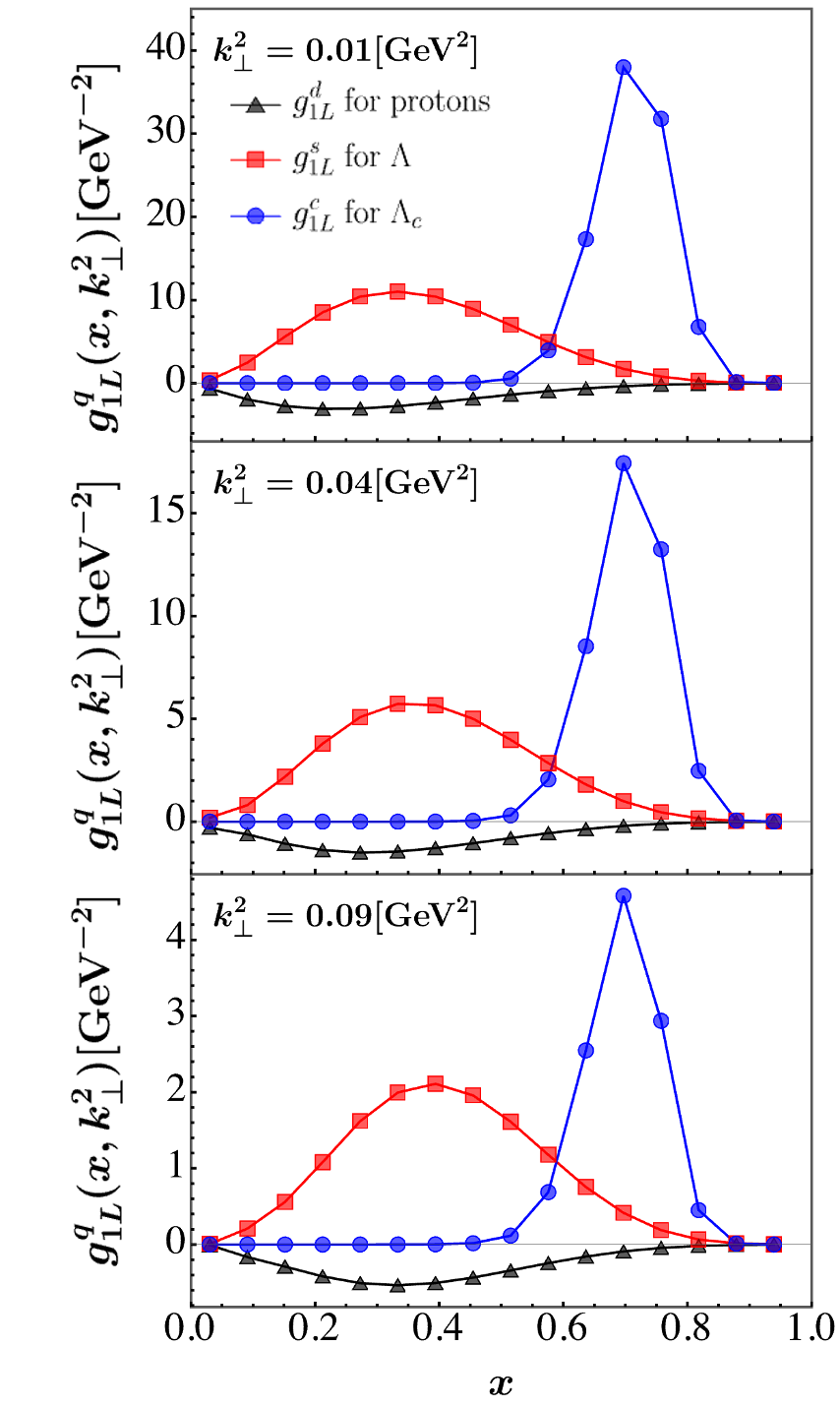}
    \includegraphics[width=0.375\textwidth]{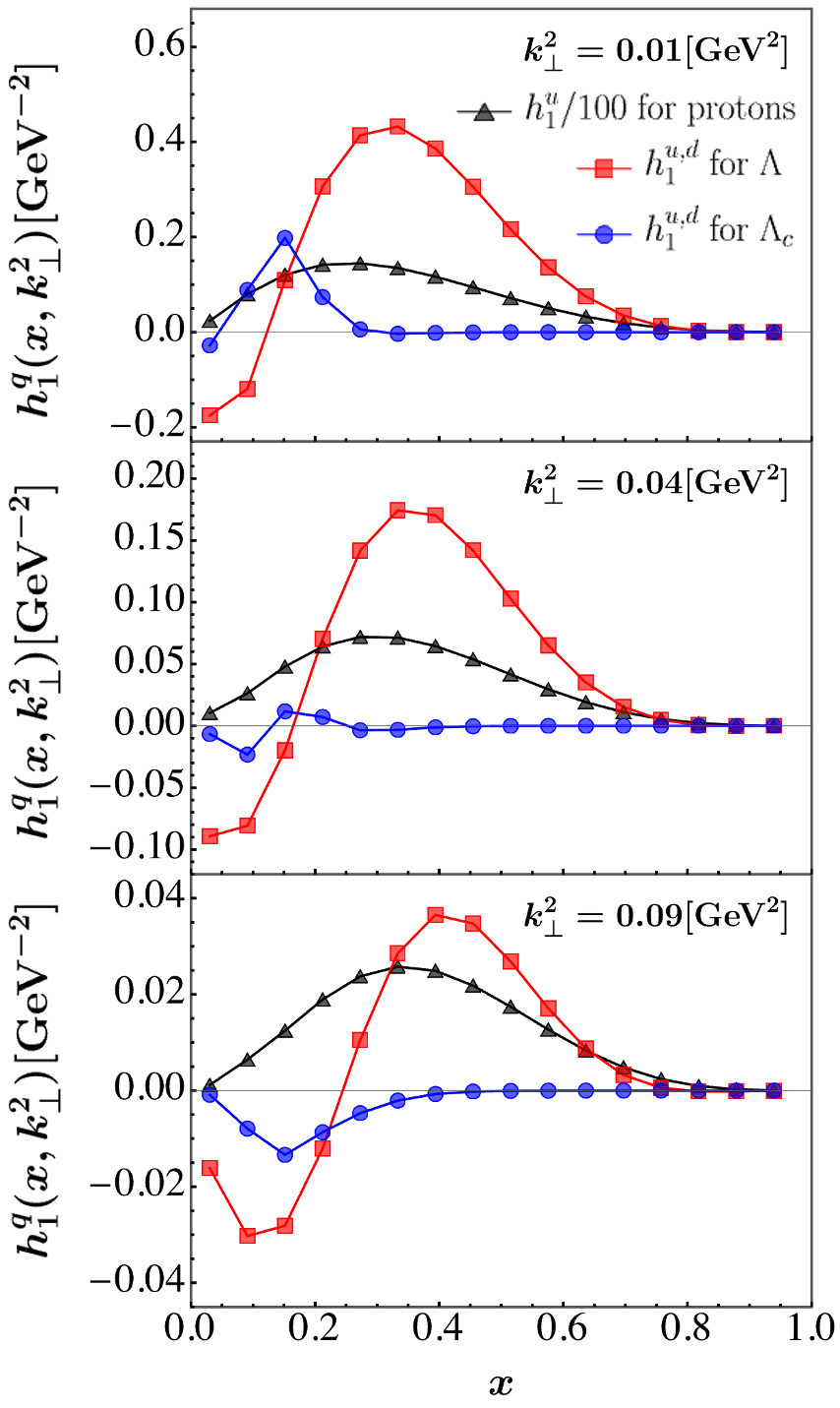}
    \includegraphics[width=0.375\textwidth]{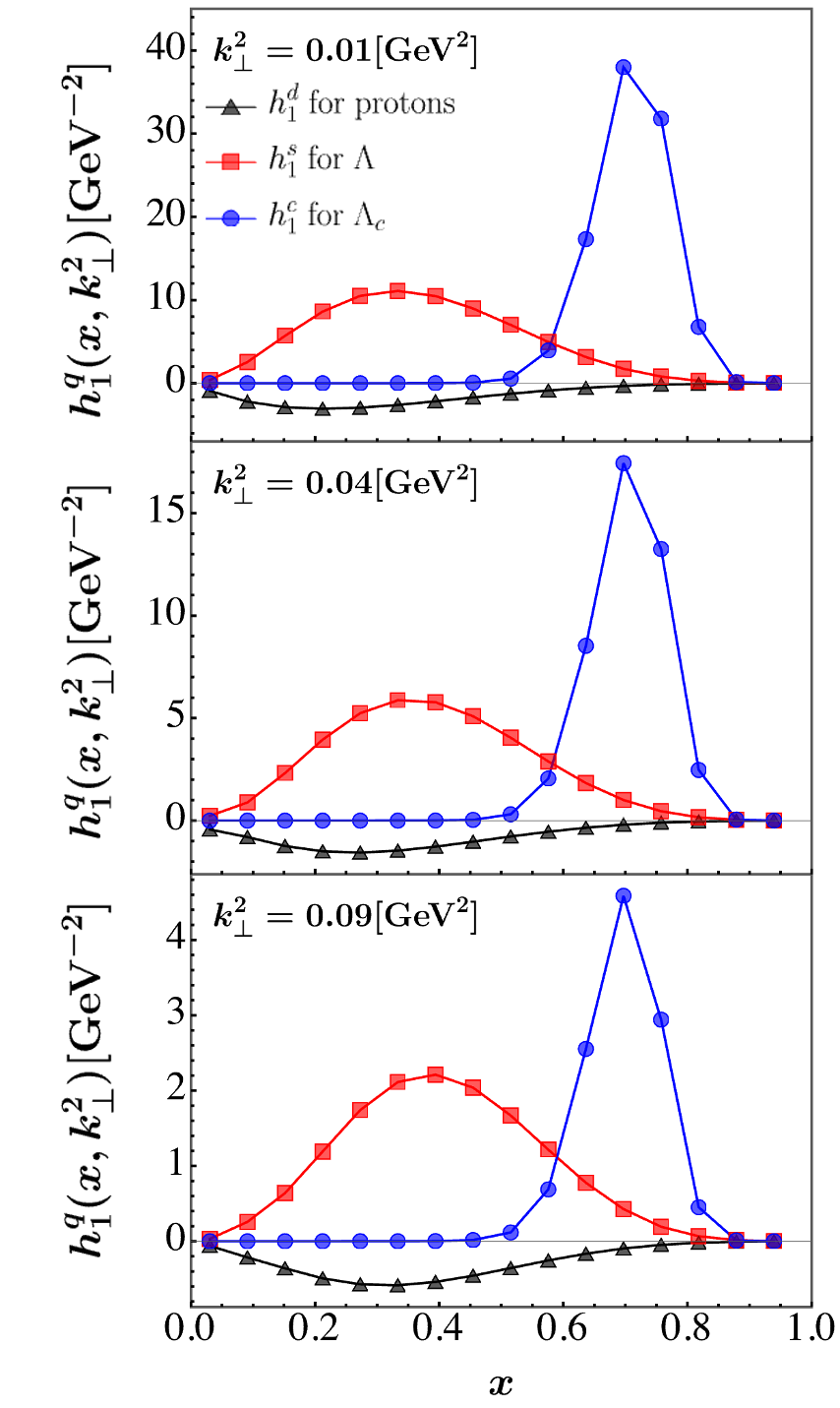}
    \caption{Two-dimensional plots for the $x$-dependence of the helicity TMD $g_{1L}$ (first row of three vertical panels) and the transversity TMD $h_1$ (second row of three vertical panels) for light quarks and `heavy' quarks inside protons, $\Lambda$, and $\Lambda_c$. Light (heavy) quarks are compared in the left (right) column. For the proton, `light' refers to the $u$ quarks and `heavy' refers the $d$ quark. For $\Lambda$, the $s$ quark plays the role of the `heavy' quark.}
    \label{comparison_g1L_and_h1_x_direction}
\end{figure*}

Figure~\ref{comparison_g1L_and_h1_x_direction} shows the $x$-dependence of the helicity TMD $g_{1L}$ and the transversity TMD $h_1$ at different $k_\perp^2$. The TMD $g_{1L}$ ($h_1$) describes the difference in the distribution partons with opposite longitudinal (transverse) polarization in longitudinally (transversely) polarized baryons, which reflects the internal spin structure of baryons~\cite{Tangerman:1994eh,Barone:2001sp,Bacchetta:1999kz,Pasquini:2012jm}. 

In Sec.~\ref{SpinStructure}, we discussed the picture for the spin structure of $\Lambda$, $\Lambda_c$, and the proton in the quark model. The $u$ and $d$ quarks have no contribution to the spin of the $\Lambda$ and $\Lambda_c$ baryons in S waves, while the $u$ quarks primarily contribute to the spin of the proton. Although our BLFQ results contain S-wave, P-wave, and D-wave contributions, the contribution from the S wave strongly dominates over the P-wave and D-wave contributions. In the first column of Fig.~\ref{comparison_g1L_and_h1_x_direction}, the amplitudes of the $g^u_{1L}$ and the $h^{u}_1$ for the $\Lambda$ baryon, and the $\Lambda_c$ baryon are not zero, because the presence of P-wave and D-wave components results in light quarks with the spin parallel to $\Lambda$ and $\Lambda_c$. However, the $g^u_{1L}$ and $h^{u}_1$ are much smaller than those in the proton, which means light quarks inside the $\Lambda$ and $\Lambda_c$ baryons are less polarized compared to the proton. 

In Sec.~\ref{SpinStructure}, another implication of the simple picture is that heavy quarks primarily contribute to the spin of $\Lambda$ and $\Lambda_c$ in the S wave, while the $d$ quark has a small but negative contribution to the spin of the proton. In the second column of Fig.~\ref{comparison_g1L_and_h1_x_direction}, $g_{1L}$ and $h_1$ for heavy quarks inside the $\Lambda$ and $\Lambda_c$ baryons are positive, but those for the $d$ quark in the proton are negative. This shows that the spin of $s$ ($c$) quark is primarily parallel to the $\Lambda$ ($\Lambda_c$) baryon, while the spin of $d$ quark is primarily antiparallel to the proton.

\section{summary\label{Sec5}}
We introduced the basis light-front quantization (BLFQ) approach, a method of obtaining the bound state wave functions by solving the eigenvalue equation of the light-front Hamiltonian. Under the BLFQ framework, we further introduced the model Hamiltonian to solve the structures of strange and charmed baryon systems. Then we obtained the LFWFs of $\Lambda$ and $\Lambda_c$ by diagonalization of the light-front Hamiltonian in an efficient basis representation.

Employing the LFWFs, we calculated the T-even twist-2 TMDs of the strange and charmed baryons. The TMDs satisfy all established inequalities and are independent of each other. Our results show that heavier quarks carry a larger longitudinal momentum fraction. Further, using the TMDs, we obtained the spin densities of valence quarks inside the $\Lambda$ and $\Lambda_c$ baryons. We discussed the probabilistic interpretations of twist-2 TMDs in connection with the quark model, and compared the TMDs of the $\Lambda$ and $\Lambda_c$ baryons to the proton. The comparison illustrates the differences in the baryon structures caused by different flavor symmetries and masses of the valence quarks, as expected from the quark model. 

The main purpose of this study is to understand the structure of the baryons in the valence Fock sector. Considering the difficulty in accessing parton structures of heavy baryons experimentally, future studies will focus on particle structures that can be accessed experimentally, such as the T-odd TMDs of the proton and the pion. 

\section*{Acknowledgements}
We thank Xiang Liu, Jiangshan Lan and Zhe Liu for many helpful discussions. 
C. M. is supported by new faculty start up funding the Institute of Modern Physics, Chinese Academy of Sciences, Grants No. E129952YR0. C. M. also thanks the Chinese Academy of Sciences Presidents International Fellowship Initiative for the support
via Grants No. 2021PM0023. X. Z. is supported by new faculty startup funding by the Institute of Modern Physics, Chinese Academy of Sciences, by Key Research Program of Frontier Sciences, Chinese Academy of Sciences, Grant No. ZDB-SLY-7020, by the Natural Science Foundation of Gansu Province, China, Grant No. 20JR10RA067, by the Foundation for Key Talents of Gansu Province, by the Central Funds Guiding the Local Science and Technology Development of Gansu Province, Grant No. 22ZY1QA006, and by the Strategic Priority Research Program of the Chinese Academy of Sciences, Grant No. XDB34000000. J. P. V. is supported by the Department of Energy under Grant No. DE-FG02-87ER40371.
A portion of the computational resources were also provided by Gansu Computing Center.

\bibliography{TMDsRef}

\end{document}